\documentclass[12pt]{article}

\usepackage{arxiv_package}

\usepackage{mathrsfs}

\usepackage{enumitem}

\usepackage[utf8]{inputenc} % allow ut-8 input
\usepackage[T1]{fontenc}    % use 8-bit T1 fonts
\usepackage{url}            % simple URL typesetting
\usepackage{hyperref}       % hyperlinks
\usepackage{booktabs}       % professional-quality tables
\usepackage{amsfonts}       % blackboard math symbols
\usepackage{nicefrac}       % compact symbols for 1/2, etc.
\usepackage{microtype}      % microtypography
\usepackage{xcolor}
\definecolor{reviewgreen}{rgb}{0,0.5,0}
\DeclareRobustCommand{\reviewchange}[2]{{\color{reviewgreen}#1}%
  \if\relax\detokenize{#2}\relax\else\ {\color{gray}[was: #2]}\fi}
\DeclareRobustCommand{\reviewdelete}[1]{{\color{gray}[deleted: #1]}}
\DeclareRobustCommand{\reviewnote}[1]{{\color{reviewgreen}[#1]}}
\usepackage{bm}
\usepackage{amsthm}
\usepackage{comment}
\usepackage{cleveref}
\usepackage{amsmath}
\usepackage{multicol, multirow, colortbl}
\usepackage{pifont}
\usepackage{caption}
\usepackage{subcaption}
\usepackage{tabularx}
\usepackage{multirow}
\usepackage{makecell}
\usepackage{wrapfig}
\usepackage{float}
\usepackage{graphicx}

\usepackage{appendix}

\usepackage{bbm}
\usepackage{dsfont}

\DeclareMathAlphabet{\mathbbold}{U}{bbold}{m}{n}
\newcommand*{\one}{\mathbbold{1}}

\hypersetup{
    colorlinks=true,
    citecolor = blue,
    linkcolor=blue
}

\makeatletter
\let\original@footnotemark\footnotemark
\newcommand{\align@footnotemark}{%
  \ifmeasuring@
    \chardef\@tempfn=\value{footnote}%
    \original@footnotemark
    \setcounter{footnote}{\@tempfn}%
  \else
    \iffirstchoice@
      \original@footnotemark
    \fi
  \fi}
\pretocmd{\start@align}{\let\footnotemark\align@footnotemark}{}{}
\makeatother

\def\independenT#1#2{\mathrel{\rlap{$#1#2$}\mkern2mu{#1#2}}}

\usepackage{mathtools}

\newcolumntype{C}{>{\centering\arraybackslash}X}

\usepackage{mathtools}

\def\E{\mathbb{E}}

\def\P{\mathbb{P}}

\makeatletter
\@ifundefined{theorem}{\newtheorem{theorem}{Theorem}[section]}{}
\@ifundefined{corollary}{}{}
\@ifundefined{proposition}{\newtheorem{proposition}{Proposition}[section]}{}
\@ifundefined{assumption}{\newtheorem{assumption}{Assumption}[section]}{}
\@ifundefined{definition}{\newtheorem{definition}{Definition}[section]}{}
\@ifundefined{example}{}{}
\makeatother
\usepackage[algoruled]{algorithm2e}

\DeclareMathOperator{\var}{Var} % if you really want lowercase \var too
\def\1{\mathbbm{1}}

\usepackage[normalem]{ulem}
\usepackage{cancel}

\title{The Reach of Treatment Effects: Randomization Inference for Zero-Effect Regions}

		\author[]{Pengyu Li\footnote{Email: \texttt{pengyuli@u.nus.edu}} \hspace{15pt} Yao Zhang\footnote{Department of Statistics and Data Science, National University of Singapore}}

\begin{document}

\maketitle

\begin{abstract}
A treatment may affect some individuals while leaving others
unaffected. We study the reach of treatment effects by asking
how large an unaffected covariate region could be.
We develop efficient local randomization tests over
nearest-neighbour regions to construct a finite-sample valid upper
confidence bound for the size of the largest zero-effect region.
A small bound provides evidence that treatment effects extend
broadly across covariate values by ruling out large zero-effect
regions. Under a general outcome model, we characterize how
the size of the bound depends on the number of affected units,
the size of their treatment effects, and the geometry of the
true zero-effect set. We show that, when region size is measured as a sample
fraction, the excess of the upper confidence bound over
the true size decreases to zero at a minimax optimal rate
up to a logarithmic factor.
Numerical simulations assess the coverage, power
and computational efficiency of our method.
Applications to
voter-turnout, HIV clinical-trial, and AI-assisted writing data
illustrate how our method rules out large zero-effect regions
across the covariates of interest.
\end{abstract}

\section{Introduction}
\label{sec:intro}
\begin{figure}[t]
\centering
\includegraphics[width=\textwidth]{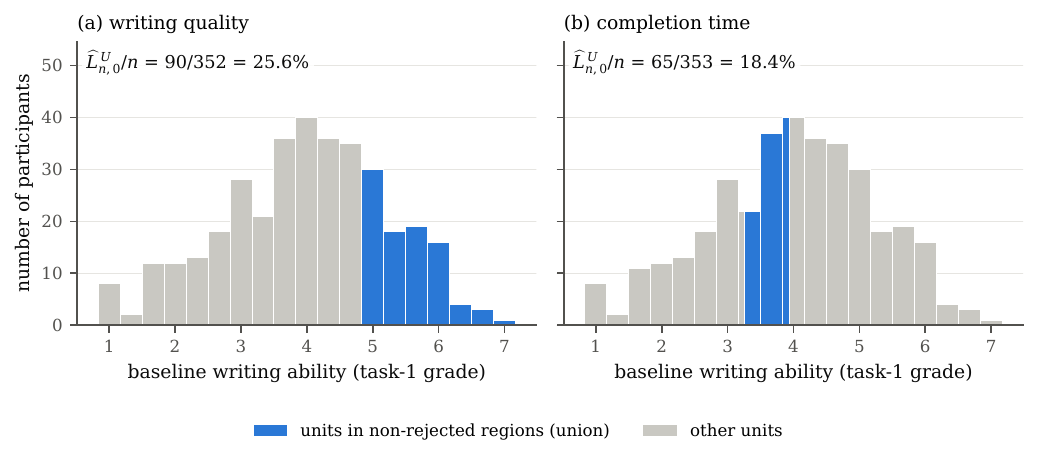}
\caption{Results from the AI-assisted writing experiment.
The $95\%$ upper confidence bounds
$\widehat L_{n,0}^{\,U}$ for the size of the largest
zero-effect region are
(a) $90$ participants ($25.6\%$) for second-task writing quality
and (b) $65$ participants ($18.4\%$) for second-task completion
time. The blue region is the union of blocks of the reported
size $\widehat L_{n,0}^{\,U}$ for which the zero-effect null
hypotheses are not rejected.}
\label{fig:real-noy}
\end{figure}

Randomized experiments are widely used to study whether a treatment
changes outcomes. Treatment effects may vary with age, baseline risk,
or prior behavior, and an average effect alone does not describe
these differences across covariate values. In a clinical trial, for
example, a drug may change outcomes for patients in some age ranges
but leave those in another range unaffected. A similar question arises
in a voter-turnout experiment: an intervention may affect participation
among people with some voting histories while having no effect among
others. In both settings, an overall treatment effect can coexist with
an unaffected covariate region.

Randomization inference methods provide valid inference on
treatment effects based on the experimental design. Fisher's randomization test
\citep{fisher1935design} has finite-sample validity for the
sharp null that treatment has a zero effect on all individual outcomes. Recent work extends randomization tests to detect a wider variety of treatment effects.
Tests of weak nulls, such as a zero sample-average effect, can be
asymptotically valid under suitable conditions
\citep{wuding2021weak,zhaoding2021covadj,cohenfogarty2022prepivoting}.
Other methods give finite-sample inference for quantiles of individual
treatment effects and for the proportion of effects above a specified
threshold \citep{caughey2023randomisation,chenli2026enhanced}.
Randomization methods also test sharp nulls within
pre-specified or selected subgroups
\citep{zhanggao2025adaptive,gao2025selective}.
A complementary question is how large a covariate region
could be in which treatment leaves every unit unaffected.
Even a statistically significant average treatment effect
can coexist with a large unaffected covariate region.  Bounding its size helps
assess how broadly treatment effects extend across
covariate values.

\textbf{Contributions.}
In this work, we develop randomization inference for the
reach of treatment effects across covariate values.
Using efficient local randomization tests, we construct
a finite-sample valid upper confidence bound for the
size of the largest zero-effect region.
We choose the pre-treatment covariates in advance and
consider nearest-neighbour regions around observed points.
A zero-effect region is one in which treatment leaves
each unit's outcome unchanged. Size is measured by the
number of units in the region, or their proportion in
the sample.
A small upper bound provides evidence against large
unaffected regions. A large bound leaves open the
possibility that treatment has no effect on a large
covariate region.

Figure~\ref{fig:real-noy} previews results from an experiment
on AI-assisted writing \citep{noy2023experimental}.
College-educated professionals completed two occupation-specific
writing tasks, with half randomly assigned access to ChatGPT
before the second task. We order participants by their
first-task grade. The $95\%$ upper confidence bounds for
the size of the largest zero-effect region are $25.6\%$
of the sample for second-task writing quality and $18.4\%$
for completion time.\footnote{The analyses use $352$ and $353$ participants,
respectively, with completion time negated. We run
the tests by permuting treatment assignments within each
analysed sample, conditional on its treated count,
assuming that sample exclusions preserve the exchangeability
of the assignments.}
For writing quality, the non-rejected regions lie among
participants with high first-task grades, leaving   
the possibility that ChatGPT does not improve writing
quality for some participants who already write well.

Like randomization tests, our upper confidence bounds have
finite-sample validity based on the experimental design.
A $95\%$ upper bound understates the true size
of the largest zero-effect region with probability at most $5\%$.
This validity requires no model assumptions about the
covariates or outcomes.
In Section~\ref{sec:sharp_limitations}, we construct these
bounds by combining local randomization tests across
locations and sizes and reporting the largest size at
which at least one local null is not rejected. Because the regions overlap, computing each local test separately repeats many
calculations. 

In Section~\ref{subsec:exact_global_test}, we propose a
solution that combines a linear statistic with shared
randomization draws, allowing cumulative sums to be
reused across tests.
For a fixed number of draws, a full one-dimensional search
has cubic cost in the sample size if each block-specific
conditional test is computed from scratch.
Our method reduces this cost to quadratic.
Under a monotonicity assumption on the computed $p$-values,
bisection reduces the number of lengths tested.
A Gaussian approximation can further reduce the number
of Monte Carlo evaluations by providing a starting point
for the search.
In Section~\ref{sec:two_dimensional_extension}, we extend
our method to two dimensions.
An exhaustive search over rectangular regions is costly, and their boundaries may poorly
describe groups on curved covariate supports.
We therefore extend the nearest-neighbour formulation,
which keeps the Monte Carlo cost quadratic in the
sample size.

A small upper confidence bound requires rejecting the
zero-effect null for every larger candidate region. In Section~\ref{sec:theory_dimensionality}, we study the
tightness of the bound by analysing the power of the local
tests across region sizes. The analysis uses a general
outcome model and allows covariates of arbitrary dimension.
A neighbourhood larger than the largest zero-effect region may
contain only a few affected units near the zero-effect boundary,
making its zero-effect null difficult to reject.
We therefore give conditions on the number of affected
units and the size of their treatment effects.
Under these conditions, the excess of our upper confidence
bound over the true size, divided by the sample size,
decreases to zero at a minimax optimal rate up to a
logarithmic factor.

Finally, in Section~\ref{sec:experiments}, we use simulations
to assess coverage and computational efficiency.
We illustrate the method using data from the
Gerber--Green--Larimer voter-turnout experiment and
the ACTG~175 HIV clinical trial.

\section{From randomization tests to zero-effect regions}
\label{sec:sharp_limitations}

\subsection{Preliminary: randomization tests for the sharp null}

Consider a completely randomized experiment with $n$ units, of
which exactly $n_1$ receive treatment. Write $[n]=\{1,\dots,n\}$.
For unit $i$, let $X_i$ denote the covariates,
$Z_i\in\{0,1\}$ the treatment
indicator, and $Y_i$ the observed outcome. We write the observation for
unit $i$ and the full data as
\[
O_i=(X_i,Y_i,Z_i),
\qquad
O_{[n]}=(X_{[n]},Y_{[n]},Z_{[n]}).
\]
Each unit has two potential outcomes, $Y_i(0)$ and $Y_i(1)$. The observed
outcome is
\[
Y_i=Y_i(Z_i)=Z_iY_i(1)+(1-Z_i)Y_i(0),
\qquad i\in[n].
\]
Fisher's global sharp null \citep{fisher1935design} is defined as
\begin{equation}
H_0:\quad Y_i(1)=Y_i(0)
\quad\text{for every }i\in[n].
\label{eq:global_sharp_null}
\end{equation}
The null is sharp because it determines every missing potential outcome:
$Y_i(z)=Y_i$ for $z\in\{0,1\}$.
Fix a test statistic $T(x_{[n]},y_{[n]},z_{[n]})$.
Under $H_0$, the observed outcomes remain unchanged under every
admissible assignment $z_{[n]}$. We write
\[
T\bigl(O_{[n]}(z_{[n]})\bigr)
:=
T\bigl(X_{[n]},Y_{[n]},z_{[n]}\bigr).
\]
This gives the exact randomization $p$-value
\[
p(O_{[n]})
=
\P_{\widetilde Z_{[n]}}
\left\{
T\bigl(O_{[n]}(\widetilde Z_{[n]})\bigr)
\ge
T(O_{[n]})
\right\},
\]
where $\widetilde Z_{[n]}$ is drawn from the known treatment assignment
distribution.

In practice, we approximate $p(O_{[n]})$ using Monte Carlo simulation.
Draw
$Z_{[n]}^{(1)},\dots,Z_{[n]}^{(M)}$
independently from the known treatment assignment distribution, and define
\[
p_M(O_{[n]})
=
\frac{
1+\sum_{m=1}^M
\one{\{
T\bigl(O_{[n]}(Z_{[n]}^{(m)})\bigr)
\ge
T(O_{[n]})\}
}
}{M+1}.
\]
Under $H_0$, the observed and Monte Carlo assignments are
exchangeable. Therefore, the observed and simulated test statistics are
also exchangeable, and $p_M(O_{[n]})$ is valid:
\begin{equation}
\P\left\{
p_M(O_{[n]})\le\alpha
\,\middle|\,
X_{[n]},Y_{[n]}(0),Y_{[n]}(1)
\right\}
\le\alpha
\qquad
\text{for all }\alpha\in[0,1].
\label{eq:validity}
\end{equation}
Here, the probability is over both the observed and Monte Carlo assignments. Thus, rejecting the sharp null when
$p_M(O_{[n]})\le\alpha$ controls the type I error at level $\alpha$.

\paragraph{What rejection and non-rejection mean.}
The alternative to the sharp null is
\[
H_1:\quad Y_i(1)\neq Y_i(0)
\quad\text{for at least one }i\in[n].
\]
Rejection therefore provides evidence that treatment changes
{\it at least one} outcome in the sample. This is a useful starting
point for a more detailed analysis of treatment effects. With this
evidence, we can move on to questions about which units are affected,
whether effects differ across units, and how broadly they extend
across covariate values. These questions require further inference:
rejection of the global null alone does not answer them. In
particular, it leaves open the possibility that a large covariate
region is unaffected.
If the null is not rejected, the chosen test does not provide enough
evidence to conclude that a treatment effect is present. This does
not establish that all effects are zero.
Other methods can help study
how average treatment effects vary across covariate values
\citep{atheyimbens2016recursive,wagerathey2018estimation}.
Their confidence intervals, however, rely on large-sample
approximations under additional assumptions. Our goal is to rule out large zero-effect regions when the data
allow, while retaining the finite-sample design-based validity. 

\subsection{Bounding the length of zero-effect regions}
\label{subsec:run_length_hypotheses}

We apply randomization tests locally to study how large an
unaffected group could be. Each local sharp null states that
treatment leaves every unit in the group unaffected. Suppose
non-rejection occurs only for small groups, while the local
sharp null is rejected for every larger candidate group.
We use this idea to construct an upper confidence bound on
the size of the largest zero-effect group.

We impose structure on the candidate groups. For simplicity, we consider
contiguous regions after ordering the units by a covariate
chosen in advance. Arbitrary groups may contain units scattered across covariate
values, making them hard to interpret. They may also contain
only weak treatment effects, making it harder to reject all
local nulls at a given size. We study the power issues formally in
Section \ref{sec:theory_dimensionality}.

Let $W_i\in\mathbb{R}$ denote the chosen covariate in $X_i$.
Without loss of generality, index units so that
$W_1 \le W_2 \le \cdots \le W_n$, with ties broken by a rule
chosen in advance. We measure a block's length by its number
of units. For $r\in[n]$, denote the collection of contiguous
blocks containing $r$ units by
\[
\mathcal B_r:=\{B_s^{(r)}:s\in[n-r+1]\}
\]
where
$B_s^{(r)}:=\{s,s+1,\dots,s+r-1\}$.
For each contiguous block $B\subseteq[n]$, define
\begin{equation}
H_{0,B}:\quad Y_i(1)=Y_i(0)\ \ \text{for all } i\in B.
\label{eq:block_sharp_null}
\end{equation}
For a fixed $r$, define the union of the local nulls over all
blocks of length $r$:
\begin{equation}
H_{0}^{(r)}:\quad \exists\, B\in\mathcal{B}_r \text{ such that } H_{0,B} \text{ holds}.
\label{eq:run_length_union_null}
\end{equation}
That is, there exists a contiguous region of length $r$ in
which every unit has zero treatment effect.
We define the length of the longest zero-effect block by
\begin{equation}
L_0
:=
\max\bigl\{|B|:
B\text{ is contiguous and }Y_i(1)=Y_i(0)\text{ for every }i\in B
\bigr\},
\label{eq:L0_def}
\end{equation}
with $L_0=0$ if no nonempty zero-effect block exists. We write $L_0(W)$
when we want to emphasize the ordering covariate. Because every sub-block of a zero-effect block is also a
zero-effect block, the nulls are nested:
\[
H_0^{(r')}\Longrightarrow H_0^{(r)}
\quad\text{whenever }r'>r.
\]
By the definition of $L_0$, we have the equivalence:
\[
H_0^{(r)}\quad\Longleftrightarrow\quad L_0\ge r.
\]
For each block $B$, let $p_B$ be a valid $p$-value for the local sharp
null $H_{0,B}$. The null $H_0^{(r)}$ holds when
the local sharp null holds for at least one block of length $r$. Following the intersection--union principle
\citep{berger1982multiparameter}, we reject $H_0^{(r)}$
only when every local null for a block of length $r$
is rejected. We therefore use the maximum $p$-value
\begin{equation}
P_r
:=
\max_{B\in\mathcal B_r}p_B.
\label{eq:global_p_max}
\end{equation}
If $H_0^{(r)}$ holds, then $H_{0,B_0}$ holds for at least one
block $B_0$ of length $r$. Because $P_r$ is the maximum of the block-level
$p$-values, it is at least as large as $p_{B_0}$. Therefore, $P_r$ can be
at most $\alpha$ only if $p_{B_0}$ is also at most $\alpha$. Since
$p_{B_0}$ is valid under $H_{0,B_0}$, this event has probability at most
$\alpha$. Thus, $P_r$ is a valid $p$-value for $H_0^{(r)}$. No
independence across overlapping blocks is required for this max-$p$
validity argument.

\paragraph{Inverting over block lengths.}
We invert the tests over $r$. The $p$-values need not decrease with length, so the
first rejection does not immediately give a valid upper bound.
We report the largest non-rejected length:
\begin{equation}
\widehat L_0^{\,U}
:=\max\{r\in[n]:P_r>\alpha\}
=\max\{r\in[n]:P_r^{\,\uparrow}>\alpha\},
\label{eq:L0_upper_bound}
\end{equation}
where $P_r^{\,\uparrow}:=\max_{s\in[n]:\,s\ge r}P_s$.
We set $\widehat L_0^{\,U}=0$ if all lengths are rejected.
The equality holds even when the $p$-values are not monotone
in length. Indeed, $P_r^{\,\uparrow}>\alpha$ holds when
$P_s>\alpha$ for some $s\ge r$, so both maxima equal
the largest length with $P_s>\alpha$.

\begin{proposition}[Upper confidence bound]
\label{prop:length_bound_validity}
If each $p_B$ is valid under $H_{0,B}$ conditional on the
covariates and potential outcomes, then the bound in
\eqref{eq:L0_upper_bound} satisfies
\[
\P\left\{
\widehat L_0^{\,U}\ge L_0
\,\middle|\,
X_{[n]},Y_{[n]}(0),Y_{[n]}(1)
\right\}
\ge 1-\alpha.
\]
\end{proposition}

When the upper confidence bound $\widehat L_0^{\,U}\ge 1$, we may also display the blocks of the
reported length whose local sharp nulls are not rejected:
\[
\widehat{\mathcal B}_{\mathrm{NR}}(\alpha)
:=
\left\{
B\in\mathcal B_{\widehat L_0^{\,U}}:
p_B>\alpha
\right\},
\qquad
\widehat{\mathcal D}_\alpha
:=
\bigcup_{B\in\widehat{\mathcal B}_{\mathrm{NR}}(\alpha)}B.
\]
The set $\widehat{\mathcal D}_\alpha$ contains every ordered position that
belongs to at least one non-rejected block of length
$\widehat L_0^{\,U}$. Because several non-rejected blocks may overlap,
their union can be longer than $\widehat L_0^{\,U}$ and need not itself
form a single block.

\paragraph{A nearest-neighbour formulation.}
\label{par:nearest_neighbour_formulation}
We give an equivalent formulation of the method above
using nearest neighbours in one dimension. In Section~\ref{sec:two_dimensional_extension}, we use this
alternative formulation to extend the method to two covariates.

Since the units are ordered by $W$, define the rank
distance between units $i$ and $j$ as $d(i,j):=|i-j|/n$.
Let $N_i^{(r)}$ contain the $r$ units nearest to $i$, including
$i$ itself, with distance ties broken in favour of the smaller
rank. Each $N_i^{(r)}$ is a contiguous block of length $r$.
Conversely, every block $B_s^{(r)}$ is the neighbourhood
$N_{s+\lfloor r/2\rfloor}^{(r)}$. Thus,
\begin{equation}
\{N_i^{(r)}:i\in[n]\}=\mathcal B_r,
\qquad
P_r=\max_{i\in[n]}p_{N_i^{(r)}}.
\label{eq:1d_neighbour_blocks}
\end{equation}
Different units $i$ may have the same set of $r$ nearest
neighbours. Counting the same block more than once does not
change the maximum.  The null hypotheses in
\eqref{eq:run_length_union_null}, the target $L_0$ in
\eqref{eq:L0_def}, and the upper confidence bound in
\eqref{eq:L0_upper_bound} are therefore identical under the
two formulations. We next introduce the methods to compute this bound.

\section{Fast computation}
\label{subsec:exact_global_test}

The upper confidence bound in \eqref{eq:L0_upper_bound} requires
local randomization tests across many overlapping blocks. Computing each
block-level $p$-value by a separate conditional permutation
test is costly. We combine shared randomization draws with a
linear statistic to reduce this cost while retaining
finite-sample validity.

\subsection{Conditional permutation is costly}
Fix a length $r$ and a block $B\in\mathcal B_r$. Let
$K_B=\sum_{i\in B}Z_i$ be the observed number of treated units
in $B$. Conditional on $K_B=k$, the distribution of $Z_B$ is
uniform over assignments that treat exactly $k$ units in $B$.
We draw $Z_B^{(1)},\dots,Z_B^{(M)}$ independently from this
conditional distribution and compute
\[
p_{B,M}(O_{[n]})
=
\frac{
1+\sum_{m=1}^M
\one{\{
T_B\bigl(O_{B}(Z_{B}^{(m)})\bigr)
\ge
T_B(O_{B})\}
}
}{M+1}.
\]
The statistic used below depends only on outcomes within $B$.
Under $H_{0,B}$, the observed and randomized statistics are
exchangeable given $K_B$, so $p_{B,M}$ is valid.

To obtain the bound in \eqref{eq:L0_upper_bound}, we test $H_0^{(r)}$ using the maximum of the block $p$-values over the $n-r+1$ overlapping blocks in $\mathcal B_r$.
Suppose we draw $M$ assignments separately for each block. Computing the test statistic from scratch for a block of $r$ units
takes $O(r)$ time per simulated assignment, so the cost of testing
all blocks of length $r$ is
\begin{equation}
C_{\mathrm{conditional}}(r)
=O\!\left(M(n-r+1)r\right).
\label{eq:conditional_cost_r}
\end{equation}
Searching all lengths $r\in[n]$ has total cost
\begin{equation}
C_{\mathrm{conditional}}
=O\!\left(M\sum_{r=1}^{n}(n-r+1)r\right)
=O(Mn^3).
\label{eq:conditional_cost_grid}
\end{equation}
This cubic cost can make a full search impractical for experiments
with tens of thousands of units, especially if hundreds or
thousands of randomization draws are used.

\subsection{Speed-up: shared permutations and linear statistics}
\label{sec:mc_centered_linear}

Under the block null $H_{0,B}$ in \eqref{eq:block_sharp_null}, the
outcomes within $B$ are the same under any treatment assignment, regardless of how many units are treated in $B$.
We can therefore draw assignments for all $n$ units from the
experiment's joint assignment distribution and reuse them across
overlapping blocks. We call these global assignments. The number
of treated units within a block may differ across draws, as it
does under the marginal assignment distribution for $Z_B$.

We draw $Z_{[n]}^{(1)},\dots,Z_{[n]}^{(M)}$ from the distribution
of $Z_{[n]}$,  and reuse them for every block. All other inputs to the block statistic must remain fixed under $H_{0,B}$;
we use only outcomes within the block.
For each length $r\in[n]$ and block $B\in\mathcal B_r$,
we compute
\begin{equation}\label{equ:pbm}
\tilde p_{B,M}(O_{[n]})
=
\frac{1+\sum_{m=1}^M \one\{T_B(O_{B}( Z_{B}^{(m)}))\ge T_B(O_{B})\}}{M+1}.
\end{equation}
Under $H_{0,B}$, the observed and randomized statistics are
exchangeable given the covariates and potential outcomes, so $\tilde p_{B,M}(O_{[n]})$ is valid:
\[
\P\{\tilde p_{B,M}(O_{[n]})\leq\alpha
\mid X_{[n]},Y_{[n]}(0),Y_{[n]}(1)\}\leq\alpha.
\]
The block $p$-values may be dependent, even for disjoint blocks,
because they use the same global assignments. This does not
affect the validity of the maximum $p$-value
\begin{equation}\label{equ:maxpvalue}
    \tilde p_r(O_{[n]}) = \max_{B\in \mathcal B_r}\tilde p_{B,M}(O_{[n]}).
\end{equation}
Under $H_0^{(r)}$, at least one block has a true local null,
so $\tilde p_r\le\alpha$ can occur only when that block's valid
$p$-value is at most $\alpha$, an event with probability at most $\alpha$.
As in \eqref{eq:L0_upper_bound}, we report the largest
non-rejected length. We denote the bound by
\begin{equation}
\begin{aligned}
\widehat L_{n,0}^{\,U}
:=\max\{r\in[n]:\tilde p_r(O_{[n]})>\alpha\} =\max\{r\in[n]:\tilde p_r^{\,\uparrow}(O_{[n]})>\alpha\},
\end{aligned}
\label{eq:L0U_def}
\end{equation}
where $\tilde p_r^{\,\uparrow}:=\max_{s\in[n]:\,s\ge r}\tilde p_s$.
Again, the equality holds without monotonicity, and we set the bound
to zero if all lengths are rejected. The argument in
Proposition~\ref{prop:length_bound_validity} shows that
$\widehat L_{n,0}^{\,U}$ is a $(1-\alpha)$ upper confidence bound
for $L_0$. 

Define the centred covariance-based statistic
\begin{equation}
T_B(O_B)
:=\sum_{i\in B}(Z_i-\pi)(Y_i-\bar Y_B),
\label{eq:centered_stat}
\end{equation}
where $\pi=n_1/n$ is the treatment probability under complete
randomization and $\bar Y_B:=|B|^{-1}\sum_{i\in B}Y_i$.
Centering prevents the statistic from becoming large merely
because the block has a high outcome mean and more treated units.
The outcomes within $B$ and their mean remain fixed under
$H_{0,B}$, ensuring finite-sample validity.
Shared permutations allow the treated count within $B$ to vary,
even when the observed block has no treated units. 
A conditional test can replace $\bar Y_B$ with predictions
from a model fitted to all covariates and observed outcomes
without using treatment assignments, because it keeps assignments
outside $B$ fixed. Holding such a fit fixed under shared global
permutations can invalidate the test, since outcomes outside $B$
may change under the local null.
Appendix~\ref{app:conditional_adjustment} explains this difference
and gives a counterexample.

We write $T_B$ for the observed statistic. For the $m$th global assignment, we compute
\[
T_B^{(m)}:=\sum_{i\in B}(Z_i^{(m)}-\pi)(Y_i-\bar Y_B),
\quad m\in [M].
\]
Since $\sum_{i\in B}(Y_i-\bar Y_B)=0$, expanding the statistic gives
\[
\begin{aligned}
T_B^{(m)}
=\sum_{i\in B}Z_i^{(m)}(Y_i-\bar Y_B)
-\pi\sum_{i\in B}(Y_i-\bar Y_B) =\sum_{i\in B}Z_i^{(m)}Y_i
-\bar Y_B\sum_{i\in B}Z_i^{(m)}.
\end{aligned}
\]
Each statistic depends on three block sums: the outcomes,
the treatment indicators, and their products.
For any array $f_{[n]}=(f_1,\dots,f_n)$ and set of units
$B\subseteq[n]$, write
\[
Q_B(f_{[n]}):=\sum_{i\in B}f_i.
\]
We use $1_{[n]}$ for the vector of ones and $\odot$ for
elementwise multiplication. In particular,
$Q_B(1_{[n]})=|B|$.
We compute these block sums from the cumulative-sum array
\[
F_j(f_{[n]}):=\sum_{i=1}^j f_i,
\qquad j\in[n],
\]
with $F_0(f_{[n]})=0$.
For a block $B=\{s,\dots,e\}$ of length $r=e-s+1$,
\[
Q_B(f_{[n]})
=F_e(f_{[n]})-F_{s-1}(f_{[n]}).
\]
When the block moves one position to the right, we update
each sum by subtracting $f_s$ and adding $f_{e+1}$.
Applying this to the vectors $Y_{[n]}$, $Z_{[n]}^{(m)}$, and
$Z_{[n]}^{(m)}\odot Y_{[n]}$ gives the three sums needed
for the test statistic:
\begin{equation}
T_B^{(m)}
=Q_B(Z_{[n]}^{(m)}\odot Y_{[n]})
-\frac{Q_B(Y_{[n]})}{r}Q_B(Z_{[n]}^{(m)}).
\label{eq:block_stat_cumulative}
\end{equation}
The observed statistic can be computed in the same way using
the observed assignment.
The cumulative-sum array for $Y_{[n]}$ is computed once
for the whole analysis. The cumulative-sum arrays for $Z_{[n]}^{(m)}$
and $Z_{[n]}^{(m)}\odot Y_{[n]}$ are computed once per
global assignment and reused across all blocks and lengths.
Each array takes $O(n)$ time to compute, after which
each block statistic takes $O(1)$ time.

\begin{table}[t]
\centering
\small
\renewcommand{\arraystretch}{1.4}
\setlength{\tabcolsep}{4pt}
\caption{Computational costs for obtaining the upper confidence
bound in one dimension, considering all lengths $r\in[n]$.
For each descending search, $t$ counts the lengths checked
up to and including the first non-rejection; $t=n$ if all
lengths are rejected. The preliminary Gaussian search evaluates
the approximation at $O(\log n)$ lengths. Here, $k$ counts the
block lengths at which all block tests are evaluated using
Monte Carlo draws.
If the two adjacent endpoints found using the Gaussian
approximation satisfy \eqref{eq:mc_bracket_conditions},
then $k=2$.
Monotonicity refers to \eqref{eq:mc_monotonicity}.
}
\label{tab:computational_costs}

\begin{tabular}{@{}
    p{0.19\linewidth}
    p{0.30\linewidth}
    p{0.18\linewidth}
    p{0.22\linewidth}
@{}}
\hline
\textbf{Permutation} & \textbf{Search}
& \textbf{Assumption} & \textbf{Cost} \\
\hline

\multirow{2}{*}{Conditional}
& Full search
& None
& $O(Mn^3)$ \\

& Descending search
& None
& $O(Mnt^2)$ \\

\hline

\multirow{4}{*}{Shared}
& Full search
& None
& $O(Mn^2)$ \\

& Descending search
& None
& $O(Mn+Mt^2)$ \\

& Bisection
& Monotonicity
& $O(Mn\log n)$ \\

& Bisection with warm-start
& Monotonicity
& $O(n\log n+Mnk)$ \\

\hline
\end{tabular}
\end{table}

The shared method draws $M$ global assignments once and reuses
them across all blocks and lengths. Computing the cumulative
sums takes $O(n)$ time per draw. Each block's Monte Carlo count
is the number of randomized statistics at least as large as its
observed statistic. At a fixed length, computing all block
statistics and updating these counts also takes $O(n)$ time
per draw, giving
$C_{\mathrm{shared}}(r)=O(Mn)$. The observed block statistics
take an additional $O(n)$ time at each length.
Compared with the conditional cost in
\eqref{eq:conditional_cost_r}, the shared method avoids drawing
assignments separately for each block and summing all $r$ terms
again for every statistic. Over all lengths $r\in[n]$, the total
cost is $C_{\mathrm{shared}}=O(Mn^2)$, compared with $O(Mn^3)$
for the conditional approach in \eqref{eq:conditional_cost_grid}.
Reducing the cost to quadratic makes it more practical to analyze larger samples, search more lengths, or use more randomization draws.

To compute the bound, we can check lengths from largest to
smallest and stop at the first non-rejection. Every larger
length has then been rejected, so this is the largest
non-rejected length in \eqref{eq:L0U_def}. If all lengths
are rejected, we report zero. This search does not require
monotone $p$-values.
Let $t$ be the number of lengths checked.
Computing the cumulative sums for the $M$ shared draws
takes $O(Mn)$ time.
At lengths $n,n-1,\dots,n-t+1$, there are
$1,2,\dots,t$ blocks, respectively.
Testing these blocks takes $O(Mt^2)$ time, giving a total
cost of $O(Mn+Mt^2)$ for the shared descending search.
The same stopping rule applies to the conditional tests.
If that search checks $t$ lengths, its cost is
$O\!\left(M\sum_{r=n-t+1}^{n}(n-r+1)r\right)
=O(Mnt^2)$.
The two methods may stop at different lengths, so $t$ is
counted separately for each method.

\subsection{Faster search under monotonicity}
\label{subsec:monotone_search}

The null hypotheses are nested: a zero-effect block of length
$r'$ contains a zero-effect block of every smaller length $r$.
Thus, the null becomes more restrictive as the length increases.
This nesting structure motivates the following monotonicity assumption,
which is not required by the search in the previous subsection:
\begin{equation}
\tilde p_{r'}\le\tilde p_r
\qquad
\text{for all }r,r'\in[n]\text{ with }r<r'.
\label{eq:mc_monotonicity}
\end{equation}
This is an additional assumption on the computed Monte Carlo
$p$-values; nesting of the nulls alone does not imply it.
For the validity statement below, we need to assume that it holds
almost surely over the observed and simulated assignments.

Under \eqref{eq:mc_monotonicity}, once a length is rejected,
every larger length is also rejected.
We can therefore use bisection to find the largest non-rejected
length. Once two adjacent lengths have a non-rejection
at the lower endpoint and a rejection at the upper endpoint,
we report the lower endpoint without checking larger lengths.
This returns the same bound as \eqref{eq:L0U_def}, so its
finite-sample validity is preserved under the monotonicity
assumption. If length $1$ is rejected, we report zero;
if length $n$ is not rejected, we report $n$.
The search evaluates
$O(\log n)$ lengths, giving
$C_{\mathrm{direct}}=O(Mn\log n)$.

A Gaussian approximation of the $p$-value can be used to first suggest two lengths between which the largest non-rejected length may lie.
We call this pair an initial bracket and this preliminary
search a Gaussian warm-start. The construction is given in
Appendix~\ref{app:gaussian_warmstart} and costs $O(n\log n)$.
If checking and refining the bracket requires Monte Carlo
tests at $k$ block lengths, the total cost is
\[
C_{\mathrm{warm}}=O(n\log n+Mnk).
\]
If the two adjacent endpoints found using the Gaussian
approximation satisfy \eqref{eq:mc_bracket_conditions},
then $k=2$ and $C_{\mathrm{warm}}=O(n\log n+Mn).$
The Monte Carlo step then tests all blocks at two lengths.
The case $k=2$ is more likely when the Gaussian approximation
is accurate near the rejection threshold and its $p$-values
at the two endpoints lie far enough above and below $\alpha$
that approximation error and Monte Carlo variation do not
change either decision. A larger $M$ reduces the Monte Carlo variation,
although additional checks may still be needed when an
endpoint $p$-value is close to $\alpha$.

Without monotonicity, the endpoint checks do not rule out
a non-rejection at a larger length. This warm-start must
still evaluate the same lengths required by the shared
descending search, while adding the Gaussian calculations
and possibly extra endpoint checks. It does not
reduce the Monte Carlo work needed for the final bound.

\textbf{Summary.}
Table~\ref{tab:computational_costs} summarizes the computational
costs. Our main computational gain comes from reusing
randomization draws and cumulative sums across regions
(\Cref{sec:mc_centered_linear}). Compared with computing each
conditional block test from scratch, this reduces the
worst-case cost of the full search from cubic to quadratic
in the sample size, without requiring monotone $p$-values.
In practice, we recommend the shared descending search when
monotonicity is not guaranteed. It computes the bound over
all lengths $r\in[n]$ and stops at the first non-rejection. Under
\eqref{eq:mc_monotonicity}, shared bisection reduces the number
of lengths evaluated to $O(\log n)$. When $M$ is large, an
accurate Gaussian approximation can further reduce the
Monte Carlo work.

\section{Extension to two dimensions}
\label{sec:two_dimensional_extension}

The two equivalent formulations in Section~\ref{sec:sharp_limitations}
suggest two extensions of our method to the setting with two
pre-treatment covariates, $W_i=(W_{i1},W_{i2})\in\mathbb R^2$,
chosen in advance. 
We show that following the original block formulation leads us
to consider rectangular regions, but computing the upper confidence bound for the size of the largest zero-effect rectangle by
exhaustive search becomes computationally prohibitive.
We instead
generalize the alternative formulation to construct nearest-neighbour
regions around each observed point, and offer an efficient
approach for computation.

\subsection{Rectangular formulation}
\label{subsec:rectangular_formulation_2d}

Let $g_{i1}$ and $g_{i2}$ be the ranks of unit $i$ in the two
coordinates, with ties broken by a rule chosen in advance.
The original one-dimensional regions are rank intervals.
We take the Cartesian
product of an interval in each coordinate:
\begin{equation}
R=\{s_1,\dots,e_1\}\times\{s_2,\dots,e_2\},
\qquad 1\le s_\ell\le e_\ell\le n,\quad \ell=1,2.
\label{eq:rectangular_regions_2d}
\end{equation}
The units in this rectangle form the set
\[
B_R:=\{i\in[n]:s_1\le g_{i1}\le e_1,\,
                       s_2\le g_{i2}\le e_2\}.
\]
Let $\mathcal R$ contain all the non-empty rectangles with
$B_R\ne\emptyset$. 
As in one dimension, the region size is $|B_R|$,
the number of units it contains.

For each $R\in\mathcal R$, use the local sharp null
$H_{0,B_R}$ in \eqref{eq:block_sharp_null}. The inferential target is the
largest number of units in a zero-effect rectangle:
\begin{equation}
L_{0,\mathrm{rect}}
:=\max\bigl\{|B_R|:R\in\mathcal R,\ H_{0,B_R}\text{ holds}\bigr\}.
\label{eq:rectangular_target_2d}
\end{equation}
At size $r$, the null $L_{0,\mathrm{rect}}\ge r$ states that
at least one rectangle containing $r$ or more units has zero
effect on every unit. Let
$\tilde p_{R,M}:=\tilde p_{B_R,M}(O_{[n]})$ be the local
shared-permutation $p$-value from \eqref{equ:pbm}. We test this
null using
$\tilde p_r^{\mathrm{rect}}
:=\max_{\substack{R\in\mathcal R\\|B_R|\ge r}}
\tilde p_{R,M}$.
The same max-$p$ argument applies. The corresponding bound is
\[
\widehat L_{0,\mathrm{rect}}^{\,U}
:=\max\{r\in[n]:\tilde p_r^{\mathrm{rect}}>\alpha\}=\max\bigl\{|B_R|:R\in\mathcal R,\ \tilde p_{R,M}>\alpha\bigr\},
\]
with an empty maximum set to zero.
The maximum includes all rectangles containing at least $r$
units. Thus, $\tilde p_r^{\mathrm{rect}}$ is non-increasing:
increasing $r$ only removes rectangles from the maximum.
The argument in Proposition~\ref{prop:length_bound_validity}
shows that $\widehat L_{0,\mathrm{rect}}^{\,U}$ is a finite-sample valid
$(1-\alpha)$ upper confidence bound for $L_{0,\mathrm{rect}}$.

\paragraph{Computational cost of $O(Mn^4)$.}
The cumulative-sum idea in Section~\ref{sec:mc_centered_linear}
can be applied to rectangles. For a rectangle $R$, write
$Q_R(f_{[n]}):=Q_{B_R}(f_{[n]})$.
We compute these sums from a two-dimensional cumulative-sum
array whose entry $F_{ab}(f_{[n]})$ sums over units with rank
pairs in $[a]\times[b]$:
\[
F_{ab}(f_{[n]})
:=\sum_{i:g_{i1}\le a,\,g_{i2}\le b}f_i,
\qquad a,b\in[n].
\]
Set entries with either index equal to zero to zero.
For the rectangle in \eqref{eq:rectangular_regions_2d}, start
with $F_{e_1,e_2}(f_{[n]})$ and subtract the parts below the
lower rank bounds $s_1$ and $s_2$. Their overlap is subtracted
twice, so we add it back. This gives
\[
\begin{aligned}
Q_R(f_{[n]})
=F_{e_1,e_2}(f_{[n]})-F_{s_1-1,e_2}(f_{[n]}) -F_{e_1,s_2-1}(f_{[n]})
+F_{s_1-1,s_2-1}(f_{[n]}).
\end{aligned}
\]
We apply this formula to $1_{[n]}$, $Y_{[n]}$,
$Z_{[n]}^{(m)}$, and $Z_{[n]}^{(m)}\odot Y_{[n]}$.
Since $Q_R(1_{[n]})=|B_R|$, the centred statistic
in \eqref{eq:centered_stat} is
\begin{equation}
T_R^{(m)}
=Q_R(Z_{[n]}^{(m)}\odot Y_{[n]})
-\frac{Q_R(Y_{[n]})}{Q_R(1_{[n]})}Q_R(Z_{[n]}^{(m)}).
\label{eq:rectangular_stat_cumulative}
\end{equation}
The cumulative-sum arrays for $1_{[n]}$ and $Y_{[n]}$
are computed once. For each global draw $m$, we compute
the arrays for $Z_{[n]}^{(m)}$ and
$Z_{[n]}^{(m)}\odot Y_{[n]}$ and reuse them across all
rectangles.
Each array takes $O(n^2)$ time to compute, after which
each rectangle statistic takes $O(1)$ time.
However, each coordinate has $n(n+1)/2$ rank intervals,
giving $O(n^4)$ combinations of rectangle boundaries.
Computing the upper confidence bound by exhaustive search
therefore takes $O(Mn^4)$ time.

\paragraph{Geometric limitations.} Rectangles have boundaries parallel to the covariate axes,
which may be unnatural for local groups on a curved or
irregular covariate support. A rectangle covering such a
group may also include affected units outside the group.
The largest zero-effect rectangle may therefore capture
only a small part of a large zero-effect group.
A small upper confidence bound for the largest zero-effect
rectangle does not rule out large zero-effect groups
with other shapes.

\subsection{Nearest-neighbour formulation}
\label{subsec:knn_regions_2d}

We extend the nearest-neighbour formulation in
Section~\ref{sec:sharp_limitations}. For each unit $i$, we vary
the neighbourhood size from $1$ to $n$, forming regions of
every size around each observed point. Their shapes depend
on the observed covariate points and the chosen distance.

\subsubsection{Upper confidence bound}
\label{subsec:2d_nulls}

We define nearest neighbours using Euclidean distance between
the normalized ranks $(g_{i1}/n,g_{i2}/n)$.
For each unit $i$, let $j_i(\ell)$ be the index of the
$\ell$th nearest unit under this distance, counting unit $i$
itself first, so $j_i(1)=i$. We break remaining distance ties
in favour of the unit with the smaller rank in the first
coordinate.
Define the neighbourhood of size $k$ centred at $W_i$ by
\begin{equation}
N_i^{(k)}:=\{j_i(1),\dots,j_i(k)\},
\qquad i,k\in[n].
\label{eq:knn_regions_2d}
\end{equation}
Each region contains exactly $k$ units, and the regions around
a given centre $W_i$ are nested:
$N_i^{(k)}\subset N_i^{(k+1)}$ for $k<n$.
With one covariate, the candidate family is exactly the
original collection of contiguous blocks, as shown in
\eqref{eq:1d_neighbour_blocks}.

For each neighbourhood $N_i^{(k)}$, we use the local sharp null in
\eqref{eq:block_sharp_null}: treatment leaves every unit in
the region unaffected. At size $k$, we test the null that
at least one candidate neighbourhood contains only
unaffected units:
\[
H_0^{(k)}:\quad
\exists i\in[n]\text{ such that }H_{0,N_i^{(k)}}\text{ holds}.
\]
Some neighbourhoods extend farther from their centres because
the covariate points are more spread out. All neighbourhoods of size $k$ contain the same fraction $k/n$ of the sample.
Our inferential target is the size of the largest zero-effect
neighbourhood:
\begin{equation}
K_0
:=\max\left\{k\in[n]:
\exists i\in[n]\text{ such that }H_{0,N_i^{(k)}}\text{ holds}
\right\},
\label{eq:K0_knn_target}
\end{equation}
with $K_0=0$ if the set is empty.
Because the neighbourhoods around each centre are nested,
$H_0^{(k)}$ holds if and only if $K_0\ge k$.

We use the centred statistic in \eqref{eq:centered_stat}
with $B=N_i^{(k)}$. Draw $M$ independent global assignments from
the experiment's joint assignment distribution and reuse them
across all centres and sizes. Compute the local $p$-value by \eqref{equ:pbm} and define
\begin{equation}
\tilde p_{i,k}
:=\tilde p_{N_i^{(k)},M}(O_{[n]}),
\qquad
\tilde p_k:=\max_{i\in[n]}\tilde p_{i,k}.
\label{eq:knn_pvalues_2d}
\end{equation}
As in \eqref{eq:L0U_def}, we report the largest non-rejected size:
\begin{equation}
\widehat K_0^{\,U}
:=\max\{k\in[n]:\tilde p_k>\alpha\}
=\max\{k\in[n]:\tilde p_k^{\,\uparrow}>\alpha\},
\label{eq:K0U_def}
\end{equation}
where $\tilde p_k^{\,\uparrow}:=\max_{s\in[n]:\,s\ge k}\tilde p_s$,
and the bound is zero if all sizes are rejected.
By the argument in Proposition~\ref{prop:length_bound_validity},
this is a finite-sample valid $(1-\alpha)$ upper confidence
bound for $K_0$, conditional on the covariates and potential
outcomes.

As in Section~\ref{subsec:run_length_hypotheses}, we highlight
the covariate points whose indices belong to the union of
non-rejected neighbourhoods at the reported size:
\[
\widehat{\mathcal D}_\alpha^{(2d)}
:=\bigcup_{i:\,\tilde p_{i,\widehat K_0^{\,U}}>\alpha}
N_i^{(\widehat K_0^{\,U})},
\qquad \widehat K_0^{\,U}>0.
\]
We set the display to the empty set when $\widehat K_0^{\,U}=0$.
The nearest-neighbour construction extends to higher dimensions.
We focus on two dimensions for interpretability and to avoid
the difficulties of defining meaningful neighbourhoods in
high dimensions.

\subsubsection{Computation using cumulative sums}
\label{subsec:knn_computation_2d}

We compute the region sums from
Section~\ref{sec:mc_centered_linear} by taking cumulative
sums along the neighbour ordering for each centre. Write
\begin{equation}
Q_{i,k}(f_{[n]})
:=Q_{N_i^{(k)}}(f_{[n]})
=\sum_{\ell=1}^k f_{j_i(\ell)},
\qquad i,k\in[n].
\label{eq:knn_cumulative_sums_2d}
\end{equation}
These sums form a cumulative-sum array with one row for each
centre $i$ and one column for each neighbourhood size $k$.
For each centre $i$, we start with $Q_{i,0}(f_{[n]})=0$
and compute the sums for $k=1,\dots,n$ by adding
$f_{j_i(k)}$ to the sum for size $k-1$.
Computing the sums for $n$ centres takes $O(n^2)$ time.

We form these arrays for $Y_{[n]}$, $Z_{[n]}^{(m)}$, and
$Z_{[n]}^{(m)}\odot Y_{[n]}$.
Since $Q_{i,k}(1_{[n]})=k$, the outcome mean in
$N_i^{(k)}$ is $Q_{i,k}(Y_{[n]})/k$.
The statistic in \eqref{eq:block_stat_cumulative} becomes
\begin{equation}
T_{i,k}^{(m)}
=Q_{i,k}(Z_{[n]}^{(m)}\odot Y_{[n]})
-\frac{Q_{i,k}(Y_{[n]})}{k}Q_{i,k}(Z_{[n]}^{(m)}).
\label{eq:knn_stat_cumulative}
\end{equation}
For each global
draw $m$, we compute the arrays for $Z_{[n]}^{(m)}$ and
$Z_{[n]}^{(m)}\odot Y_{[n]}$ and reuse them across all
centres and sizes. Each statistic then takes $O(1)$ time
to compute.

Computing all pairwise distances takes $O(n^2)$ time, and
sorting the neighbour orders takes $O(n^2\log n)$ time.
These steps are done once. For each global draw, computing
the cumulative-sum arrays and updating the Monte Carlo counts
across all centres and sizes takes $O(n^2)$ time.
The total cost of the full search is $O(n^2\log n+Mn^2).$
The Monte Carlo cost is of the same order as the full one-dimensional search.

\paragraph{Descending search.}
As in Section~\ref{sec:mc_centered_linear}, we can check sizes
from largest to smallest and stop at the first non-rejection.
This computes \eqref{eq:K0U_def} without requiring monotone
$p$-values. The nesting of the nulls alone does not justify
bisection; see Section~\ref{subsec:monotone_search}.

At $k=n$, all neighbourhoods equal
$[n]$, so their sums share the same initial values.
Moving from $k$ to $k-1$ removes unit $j_i(k)$ from the
neighbourhood centred at $W_i$. We subtract this unit's
contribution from each of the three sums, for both the observed
and simulated assignments. Updating all statistics and
Monte Carlo counts at one size takes $O(Mn)$ time.
If the search checks $t$ sizes, the total cost is $O(n^2\log n+Mnt).$

\paragraph{Search under monotonicity.}
The methods in Section~\ref{subsec:monotone_search} extend
under the same assumption: the computed $p$-values
$\tilde p_k$ are non-increasing in $k$, almost surely over
the observed and simulated assignments.
Bisection then finds the bound in \eqref{eq:K0U_def}
by checking $O(\log n)$ sizes.
The Gaussian warm-start also applies, using the approximation
in Appendix~\ref{app:gaussian_warmstart} with $B=N_i^{(k)}$.
However, precomputing the neighbour cumulative-sum arrays
for all $M$ global draws already costs $O(Mn^2)$ time.
Thus, these methods do not improve the worst-case order
with this implementation.
We therefore use descending search, which updates the sums
only as needed and stops at the first non-rejection.

\section{Theory}
\label{sec:theory_dimensionality}

\subsection{Setup}

Let $W_i=(W_{i1},\dots,W_{id})\in\mathbb R^d$ be the covariates
used to define neighbourhoods. As in
Section~\ref{sec:two_dimensional_extension}, we define
$N_i^{(k)}$ as the neighbourhood of size $k$ centred at $W_i$,
using Euclidean distance between the normalized rank vectors
$(g_{i1}/n,\dots,g_{id}/n)$.

Write $\tau_i:=Y_i(1)-Y_i(0)$. The local sharp null is
$H_{0,N_i^{(k)}}:\tau_j=0$ for every $j\in N_i^{(k)}$.
The target is the largest zero-effect neighbourhood size,
\begin{equation}
K_0:=\max\left(
\left\{k\in[n]:\exists i\in[n],\,
H_{0,N_i^{(k)}}\text{ holds}\right\}\cup\{0\}
\right).
\label{eq:knn_target_d}
\end{equation}
Thus $K_0/n$ is a sample fraction. Both the candidate
neighbourhoods and this target depend on the chosen distance
and the observed covariates.

Using the cumulative sums in
\eqref{eq:knn_cumulative_sums_2d}, we write the centred
statistic in \eqref{eq:centered_stat} as
\begin{equation}
T_{i,k}(z_{[n]})
:=Q_{i,k}(z_{[n]}\odot Y_{[n]})
-\frac{Q_{i,k}(Y_{[n]})}{k}Q_{i,k}(z_{[n]}).
\label{eq:knn_stat_d}
\end{equation}
We compute the local $p$-values $\tilde p_{i,k}$ and their
maxima $\tilde p_k$ as in \eqref{eq:knn_pvalues_2d}.
As before, we report the largest non-rejected size:
\begin{equation}
\widehat K_0^{\,U}
:=\max\bigl(\{k\in[n]:\tilde p_k>\alpha\}\cup\{0\}\bigr).
\label{eq:knn_bound_d}
\end{equation}
We have the following results for the validity and computation of the confidence bound $\widehat K_0^{\,U}$. The computation includes the cost of forming the neighbour orderings.
\begin{theorem}[Validity]
\label{thm:knn_validity}
For every dimension $d\ge1$, number of simulated assignments
$M\ge1$, and  level $\alpha\in(0,1)$, the upper
confidence bound $\widehat K_0^{\,U}$ defined in
\eqref{eq:knn_bound_d} satisfies
\[
\mathbb P\!\left\{
\widehat K_0^{\,U}\ge K_0
\,\middle|\,
X_{[n]},Y_{[n]}(0),Y_{[n]}(1)
\right\}\ge1-\alpha.
\]
Thus, $[0,\widehat K_0^{\,U}/n]$ is a valid $(1-\alpha)$-confidence interval for $K_0/n$.
\end{theorem}

\begin{proposition}[Computation]
\label{prop:knn_computational_cost}
In a completely randomized experiment, the costs of computing
the bound in \eqref{eq:knn_bound_d} using cumulative sums are
\[
\begin{aligned}
\text{Full search:}\quad
& O(dn^2+n^2\log n+Mn^2),\\
\text{Descending search:}\quad
& O(dn^2+n^2\log n+Mnt),
\end{aligned}
\]
where $t$ is the number of sizes checked by descending search.
\end{proposition}

\subsection{Power analysis}
\label{subsec:d_dimensional_power}
We condition on $X_{[n]}$, so the neighbour orderings are fixed. Treatment follows complete randomization with
$\pi=n_1/n\in[\pi_{\min},1-\pi_{\min}]$, where $\pi_{\min}>0$
is fixed.

\begin{assumption}[Outcome model]
\label{ass:knn_power_model}
\[
Y_i(0)=\mu_0(X_i)+\epsilon_i,
\qquad
Y_i(1)=Y_i(0)+\tau(X_i).
\]
Conditional on $X_{[n]}$, the errors
$\epsilon_1,\dots,\epsilon_n$ are mutually independent,
mean-zero and $\sigma$-sub-Gaussian, and independent of the
treatment assignment.
For every $i$,
\[
|\mu_0(X_i)|\le\mu_{\max}<\infty,
\qquad
|\tau(X_i)|\le\tau_{\max}<\infty.
\]
\end{assumption}

Write $\tau_i:=\tau(X_i)$ and let
$s_Y:=\sigma+\mu_{\max}+\tau_{\max}$ denote the overall
outcome scale. Define the neighbourhood's contribution to
the sample average treatment effect by
\begin{equation}
\bar\tau_{i,k}:=\frac1n\sum_{j\in N_i^{(k)}}\tau_j,
\qquad i,k\in[n].
\label{eq:knn_regional_signal}
\end{equation}
This equals its local average treatment effect multiplied
by its sample fraction $k/n$.

Let $v_j=(g_{j1}/n,\dots,g_{jd}/n)$ be the normalized rank
vector of unit $j$. The distance between units $i$ and $j$
used to construct the neighbourhoods is $\|v_i-v_j\|_2$.

\begin{definition}[Zero-effect set]
\label{def:knn_zero_effect_set}
A nonempty closed set $\mathcal Z_0\subseteq[0,1]^d$
is a zero-effect set in rank space if, for every $j\in[n]$,
$\tau_j=0$ exactly when $v_j\in\mathcal Z_0$.
\end{definition}

Write $\operatorname{dist}(v,\mathcal Z_0)
:=\inf_{z\in\mathcal Z_0}\|v-z\|_2$ for the shortest
Euclidean distance from $v$ to the zero-effect set.
The target $K_0$ can also be written as
\[
K_0
=\max\left(
\left\{k\in[n]:\exists i\in[n],\
\{v_j:j\in N_i^{(k)}\}\subseteq\mathcal Z_0
\right\}\cup\{0\}
\right).
\]
The inclusion $\{v_j:j\in N_i^{(k)}\}\subseteq\mathcal Z_0$
means that every unit in the neighbourhood is unaffected.
Thus, $K_0$ is the largest number of units in any candidate
neighbourhood whose treatment effects are all zero.
We set $K_0=0$ if no such neighbourhood exists.

By definition of $K_0$, a neighbourhood larger than $K_0$ contains at least
one affected unit, but this alone may not provide enough power.
To obtain a nontrivial upper bound for $K_0$, we require
larger neighbourhoods to contain enough affected units with
large enough treatment effects. We thus make the following
assumption, which bounds treatment effects from below according
to their distance from $\mathcal Z_0$ and requires each
neighbourhood larger than $K_0$ to contain enough units
sufficiently far from $\mathcal Z_0$.

\begin{assumption}
\label{ass:knn_geometry}
For constants $\lambda_\tau,a,c>0$, $\beta\ge0$ and $\nu\ge1$,
\begin{equation}
\tau_j\ge\lambda_\tau
\operatorname{dist}(v_j,\mathcal Z_0)^\beta,
\qquad v_j\notin\mathcal Z_0.
\label{eq:knn_boundary_effects}
\end{equation}
For every $i,k\in[n]$ with $k>K_0$, write
$u_k:=(k-K_0)/n$. If $\beta>0$, suppose
\begin{equation}
\#\left\{j\in N_i^{(k)}:
\operatorname{dist}(v_j,\mathcal Z_0)\ge a u_k
\right\}
\ge cn u_k^\nu.
\label{eq:knn_geometric_count}
\end{equation}
If $\beta=0$, require each such neighbourhood to contain
at least $cn u_k^\nu$ affected units.
\end{assumption}

Assumption~\ref{ass:knn_geometry} requires each neighbourhood
of size $k>K_0$ to contain at least $cn u_k^\nu$ affected
units. Smaller $c$ or larger $\nu$ permits fewer affected
units. When $\beta>0$, these units must also lie at distance
at least $a u_k$ from $\mathcal Z_0$. Smaller $a$ allows
them to lie closer to the zero-effect set, where treatment
effects may be small. Larger $\beta$ permits effects that
grow more slowly near the boundary.

For $\beta>0$, the counted units lie at distance at least
$a u_k$ from $\mathcal Z_0$. Substituting this distance into
\eqref{eq:knn_boundary_effects} gives the lower bound
$\lambda_\tau(a u_k)^\beta$ on their treatment effects.
For $\beta=0$, the lower bound is simply $\lambda_\tau$.
Since there are at least $cn u_k^\nu$ counted units and
all other effects are nonnegative, writing
$\lambda:=\lambda_\tau c a^\beta$ and $\gamma:=\nu+\beta$ gives
\begin{equation}
\bar\tau_{i,k}
\ge\lambda\left(\frac{k-K_0}{n}\right)^\gamma,
\qquad i\in[n],\quad k>K_0.
\label{eq:knn_signal_growth}
\end{equation}
This gives a lower bound on $\bar\tau_{i,k}$ in every
neighbourhood larger than $K_0$. Smaller $\lambda$ or larger
$\gamma$ allows smaller values of $\bar\tau_{i,k}$ when
$k$ is close to $K_0$.

For a failure probability $\zeta\in(0,1)$, let
$L_n(\zeta):=\log(32n^3/\zeta)$.
The following result shows how the growth of $\bar\tau_{i,k}$
controls the gap between the bound and its target.

\begin{theorem}[Upper bound]
\label{thm:knn_power}
Suppose Assumptions~\ref{ass:knn_power_model}, ~\ref{ass:knn_geometry} and $M\ge(12/\alpha)\log(2n^2/\zeta)$. Then, with probability
at least $1-\zeta$ conditional on $X_{[n]}$,
\begin{equation}
\frac{\widehat K_0^{\,U}-K_0}{n}
\lesssim
\left(\frac{s_Y^2 K_0 L_n(\zeta)}{\lambda^2 n^2}\right)^{1/(2\gamma)}
+
\left(\frac{s_Y^2 L_n(\zeta)}{\lambda^2 n}\right)^{1/(2\gamma-1)}
+\frac1n.
\label{eq:knn_power_rate}
\end{equation}
For fixed $\zeta$, $s_Y/\lambda$, and $\gamma$, the inequality $K_0\le n$ gives,
for sufficiently large $n$,
\begin{equation}
\frac{\widehat K_0^{\,U}-K_0}{n}
\lesssim
\left(\frac{\log n}{n}\right)^{1/\{2(\nu+\beta)\}}.
\label{eq:rate_k}
\end{equation}
\end{theorem}
\Cref{thm:knn_power} controls how much the reported upper
bound $\widehat K_0^{\,U}/n$ can exceed $K_0/n$, the fraction
of the sample contained in the largest zero-effect
neighbourhood.
The quantity $s_Y=\sigma+\mu_{\max}+\tau_{\max}$ accounts
for outcome noise and the magnitudes of baseline outcomes
and treatment effects. The parameter $\lambda$, defined
before \eqref{eq:knn_signal_growth}, controls the lower bound
on the treatment effects accumulated as neighbourhoods grow
beyond $K_0$. A smaller ratio $s_Y/\lambda$ therefore gives
a tighter bound.

The geometry of the zero-effect set affects power through
this accumulation of treatment effects. An irregular or
disconnected set may leave neighbourhoods larger than $K_0$
with only a few affected units. If their effects are also
small, the local zero-effect null can be difficult to reject.
Irregularity or disconnectedness
alone does not imply lower power; what matters is how much
treatment effect these larger neighbourhoods contain.
In \eqref{eq:knn_geometric_count}, the parameter $\nu$ controls the lower
bound $cn u_k^\nu$ for the number of affected units lying
at distance at least $a u_k$ from $\mathcal Z_0$. 
Since $u_k\in(0,1]$, larger $\nu$
permits fewer such units. In \eqref{eq:knn_boundary_effects},
larger $\beta$ allows smaller treatment effects near the
boundary. Both increase $\gamma=\nu+\beta$ and give a slower convergence rate in \Cref{thm:knn_power}.

There are at most $n^2$ candidate neighbourhoods in every
dimension: one for each centre and size.
Dimension can nevertheless affect how many affected units
these neighbourhoods contain and hence the exponent $\nu$
in \eqref{eq:knn_geometric_count}.
Appendix~\ref{app:knn_dimension} illustrates this through
a ball expanding across a flat zero-effect boundary.
As the ball expands, its affected part grows more slowly
in higher dimensions. This calculation suggests
$\nu=(d+1)/2$ in Assumption~\ref{ass:knn_geometry}.
If the assumption holds with this value of $\nu$ for every
sample neighbourhood larger than $K_0$,
Theorem~\ref{thm:knn_power} gives the rate
$(\log n/n)^{1/(d+1+2\beta)}$.

Fix $d\ge1$, $\sigma>0$, $\gamma\ge1$,
and $0<\lambda\le\tau_{\max}/(16\gamma)$.
We next show that the exponent $1/(2\gamma)$ in
Theorem~\ref{thm:knn_power} cannot be improved uniformly
over the class $\mathcal F_n(\lambda,\gamma)$ defined below.
For this lower-bound argument only, we use a fixed covariate
design and a Gaussian special case of the outcome model in
Assumption~\ref{ass:knn_power_model}, specified in
Appendix~\ref{app:proof_knn_minimax_lower}.
The class $\mathcal F_n(\lambda,\gamma)$ contains all treatment effect
vectors in $[0,\tau_{\max}]^n$ satisfying
Assumption~\ref{ass:knn_geometry} with
$\lambda_\tau ca^\beta=\lambda$ and $\nu+\beta=\gamma$.
This class allows irregular treatment effects and zero-effect
sets with disconnected parts.
Write $\mathbb P_\tau$ for the law of the observed data and
any auxiliary randomness used to compute the bound.

\begin{theorem}[Lower bound]
\label{thm:knn_minimax_lower}
Fix $\alpha,\zeta\in(0,1)$ with $\alpha+\zeta<1$.
For sufficiently large $n$, any upper confidence bound $U$
satisfying
\[
\inf_{\tau\in\mathcal F_n(\lambda,\gamma)}
\mathbb P_\tau\{U\ge K_0(\tau)\}\ge1-\alpha
\]
obeys, for some $\tau\in\mathcal F_n(\lambda,\gamma)$
and with probability greater than $\zeta$,
\begin{equation}
\frac{U-K_0(\tau)}{n}\gtrsim n^{-1/(2\gamma)}.
\label{eq:knn_minimax_lower}
\end{equation}
\end{theorem}

Theorem~\ref{thm:knn_power} bounds the excess fraction
at rate $(\log n/n)^{1/(2\gamma)}$ with probability
at least $1-\zeta$.
At this probability level,
Theorem~\ref{thm:knn_minimax_lower} rules out
rates faster than $n^{-1/(2\gamma)}$ for any uniformly
valid upper confidence bound.
Thus, the rate for the excess of our upper confidence
bound over the true size is optimal up to the logarithmic
factor $(\log n)^{1/(2\gamma)}$ over the allowed covariate
designs and treatment effects in
$\mathcal F_n(\lambda,\gamma)$.

In the proof of Theorem~\ref{thm:knn_minimax_lower}, we obtain the lower bound by constructing alternative
models that assign positive effects to a few units at unknown
locations within a zero-effect neighbourhood.
These models satisfy Assumption~\ref{ass:knn_geometry}
with $\beta=0$ and $\nu=\gamma$, but their values of $K_0$
are smaller than that of the original model by order
$n^{1-1/(2\gamma)}$.
A mixture over their data distributions remains close to
the original distribution, making this difference in $K_0$
difficult to detect from the observed data.

\DeclareRobustCommand{\expupdate}[1]{{\color{blue}\hypersetup{linkcolor=blue,citecolor=blue,urlcolor=blue}#1}}
\pdfstringdefDisableCommands{\def\expupdate#1{#1}}
% Keep the figures of each subsection of Section 6 within it (defined here, not in the preamble).
\makeatletter
\providecommand\FloatBarrier{\par\begingroup\let\@elt\relax
  \edef\@tempa{\@botlist\@deferlist\@dbldeferlist}%
  \ifx\@tempa\@empty\else\clearpage\fi\endgroup}
\makeatother
% END BLUE EXPERIMENT-UPDATE MARKUP

\section{Experiments}
\label{sec:experiments}

We next test our method using synthetic and real data.
In simulations, we assess coverage, power, convergence rates,
and computational efficiency.
We then apply the method to real data to illustrate how it bounds the size of zero-effect regions across covariates of interest. Code to reproduce the experiments in this article is available at \url{https://github.com/lipengyu88/reach-of-treatment-effects}.

\begin{figure}[ht]
\centering
\begin{subfigure}[t]{\textwidth}
\centering
\includegraphics[width=\linewidth]{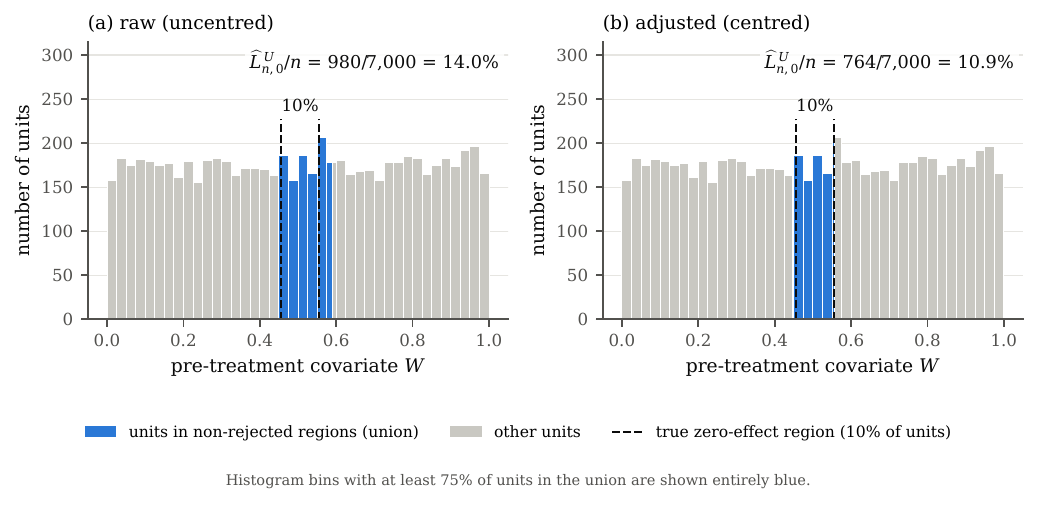}
\caption{One dimension.}
\label{fig:region-compatibility-1d}
\end{subfigure}

\medskip

\begin{subfigure}[t]{\textwidth}
\centering
\includegraphics[width=\linewidth]{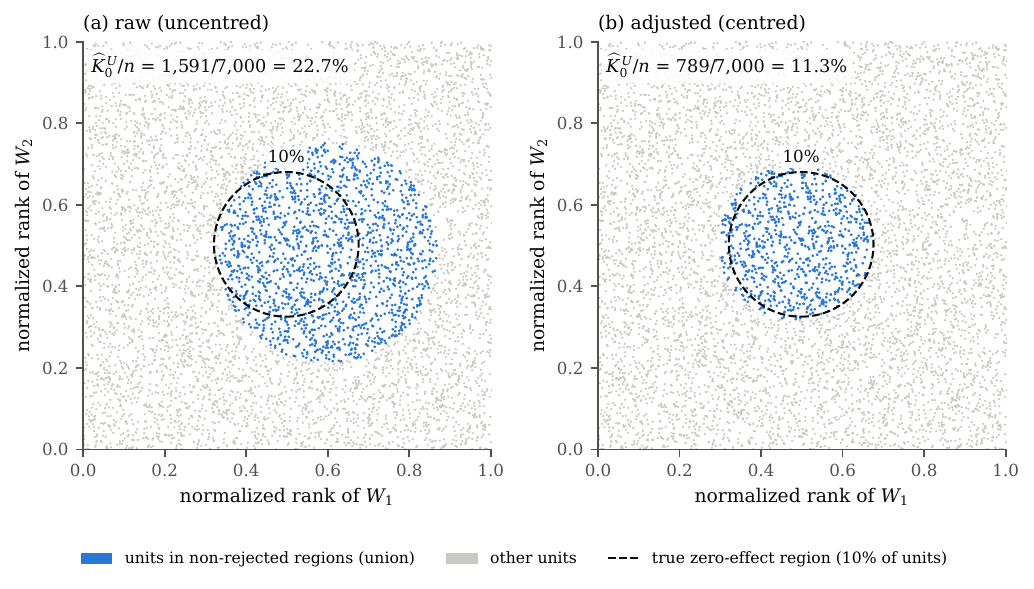}
\caption{Two dimensions.}
\label{fig:region-compatibility-2d}
\end{subfigure}

\caption{Non-rejected regions ($n=7000$).
The histogram shows the distribution of $W$ in one dimension;
the points show normalized covariate ranks in two dimensions.
Blue marks the union of non-rejected regions of the
size, $\widehat L_{n,0}^{\,U}$ in one dimension and
$\widehat K_0^{\,U}$ in two dimensions. The dashed lines mark the true zero-effect regions.
Histogram bins with at least three quarters of their units
in the union are drawn entirely blue.}
\label{fig:region-compatibility}
\end{figure}

\subsection{Synthetic data experiments}
\label{sec:exp-overview}

In one dimension, we generate
\[
W_i\sim U(0,1),\qquad
Y_i(0)=0.5+2W_i+\epsilon_i,\qquad
\epsilon_i\sim N(0,0.5^2).
\]
After ordering units by $W_i$, we set the treatment effect
to zero in a centred block of length $L_0=\lfloor 0.10n\rfloor$
and to $\delta=1$ outside it. In two dimensions, we generate
\[
W_i\sim U(0,1)^2,\qquad
Y_i(0)=0.5+2W_{i1}+2W_{i2}+\epsilon_i,
\]
with the same error distribution. We choose the unit closest
to $(0.5,0.5)$ as the centre and set the effects of its
$\lfloor 0.10n\rfloor$ nearest neighbours, including itself,
to zero. Neighbours are defined using the normalized-rank
distance in Section~\ref{subsec:2d_nulls}. All other units
have effect $\delta=1$, so $K_0=\lfloor 0.10n\rfloor$.
In both settings, $Y_i(1)=Y_i(0)+\tau_i$.

Unless stated otherwise, we use complete randomization with half the units treated,
$\alpha=0.05$, and $M=500$ shared randomization draws.
We search over all integer region sizes using descending
search and run $200$ repetitions for each sample size $n$.

For each candidate region $B$, we compare our method using
the uncentred statistic
\[
T_B^{\mathrm{raw}}=\sum_{i\in B}(Z_i-\pi)Y_i
\]
with our method using the centred statistic in
\eqref{eq:centered_stat}. In both dimensions, the comparison
uses the same outcomes, treatment assignment, candidate
regions and shared randomization draws in each repetition.
Figure~\ref{fig:region-compatibility} shows the
union of non-rejected regions for
a repetition with $n=7000$. With the uncentred
statistic, the union extends beyond the true zero-effect
region, especially in two dimensions. Centering reduces
this extension and tightens the bound from $14.0\%$ to
$10.9\%$ in one dimension and from $22.7\%$ to $11.3\%$
in two dimensions, closer to the true size of $10\%$.

\subsubsection{Coverage and size of the upper bound}
\label{sec:exp-feasible-c1c2}

Figure~\ref{fig:calibration} shows empirical coverage on the
left and mean confidence bounds divided by $n$ on the right.
Across all simulated sample sizes, empirical coverage is
at least the nominal $95\%$ level for both statistics.
Centering reduces variation in the statistic caused by differences in the number of treated units across randomization draws, making treatment effects easier to detect in smaller regions.

We can see that centering gives much smaller bounds in both dimensions,
particularly at small sample sizes. 
At $n=200$, it reduces
the mean bound divided by $n$ from $0.574$ to $0.196$
in one dimension and from $0.787$ to $0.263$ in two
dimensions. For both statistics, the mean bound approaches the true fraction $0.10$ as $n$ increases, but it does so much faster with the centred statistic.
Appendix~\ref{app:exp-rates} examines convergence under
different treatment-effect patterns and zero-effect regions.

\clearpage
\begingroup
\captionsetup{font=small,skip=4pt}

\begin{figure}[!p]
\centering
\begin{subfigure}[t]{\textwidth}
\centering
\includegraphics[
  width=0.48\linewidth,
  height=0.23\textheight,
  keepaspectratio
]{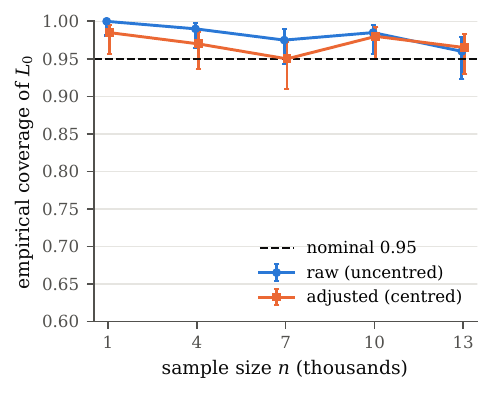}
\hspace{0.02\linewidth}%
\includegraphics[
  width=0.48\linewidth,
  height=0.23\textheight,
  keepaspectratio
]{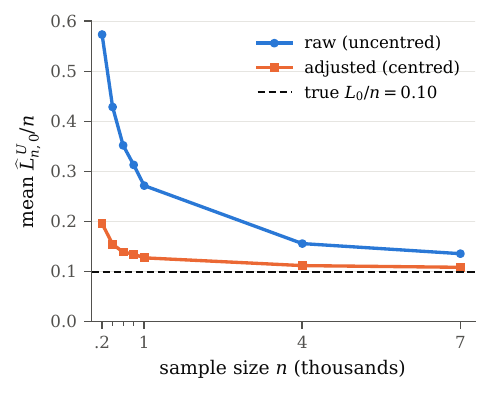}
\caption{One dimension.}
\label{fig:feasible-c1c2-coverage}
\end{subfigure}

\smallskip

\begin{subfigure}[t]{\textwidth}
\centering
\includegraphics[
  width=0.48\linewidth,
  height=0.23\textheight,
  keepaspectratio
]{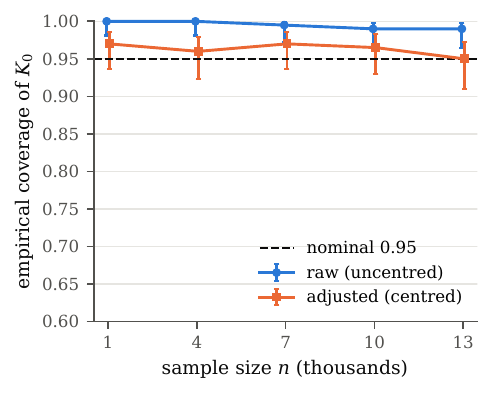}
\hspace{0.02\linewidth}%
\includegraphics[
  width=0.48\linewidth,
  height=0.23\textheight,
  keepaspectratio
]{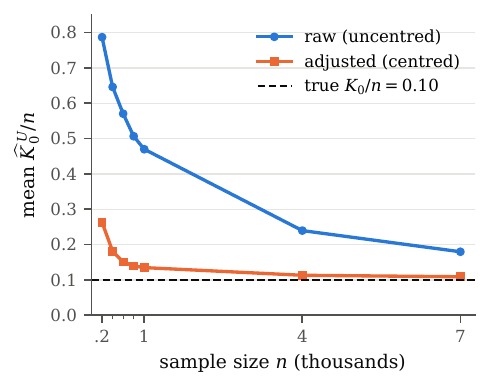}
\caption{Two dimensions.}
\label{fig:calibration-2d}
\end{subfigure}

\caption{Coverage and size of the upper bounds using the
uncentred and within-region centred statistics.
Left: empirical coverage of $L_0$ ($K_0$) in one (two)
dimensions; the dashed line marks the nominal $0.95$ level.
Right: means of the bound $\widehat L_{n,0}^{\,U}/n$
($\widehat K_0^{\,U}/n$) in one (two) dimensions;
the dashed line marks the true value $0.10$.}
\label{fig:calibration}
\end{figure}

\begin{figure}[!p]
\centering
\begin{subfigure}[t]{0.48\textwidth}
\centering
\includegraphics[
  width=\linewidth,
  height=0.20\textheight,
  keepaspectratio
]{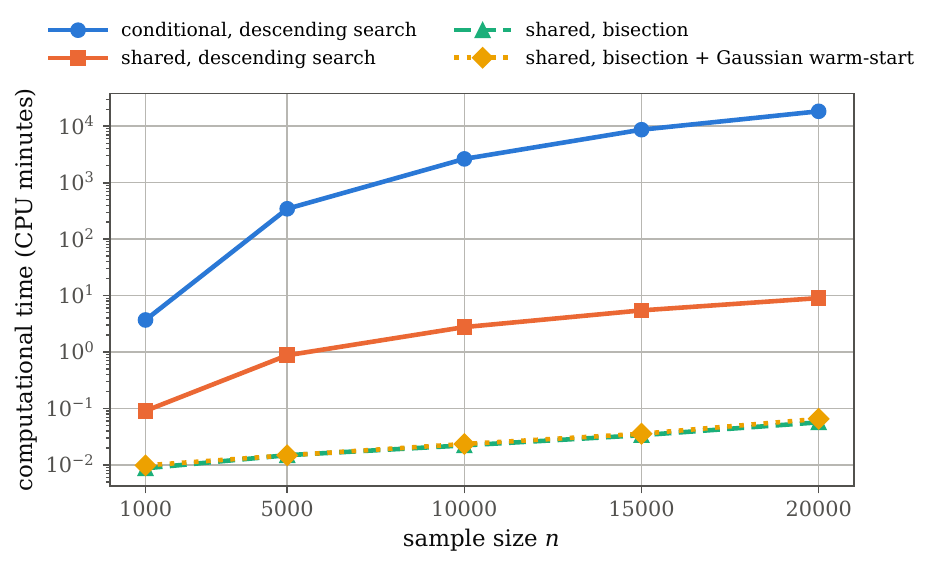}
\caption{One dimension.}
\end{subfigure}\hfill
\begin{subfigure}[t]{0.48\textwidth}
\centering
\includegraphics[
  width=\linewidth,
  height=0.20\textheight,
  keepaspectratio
]{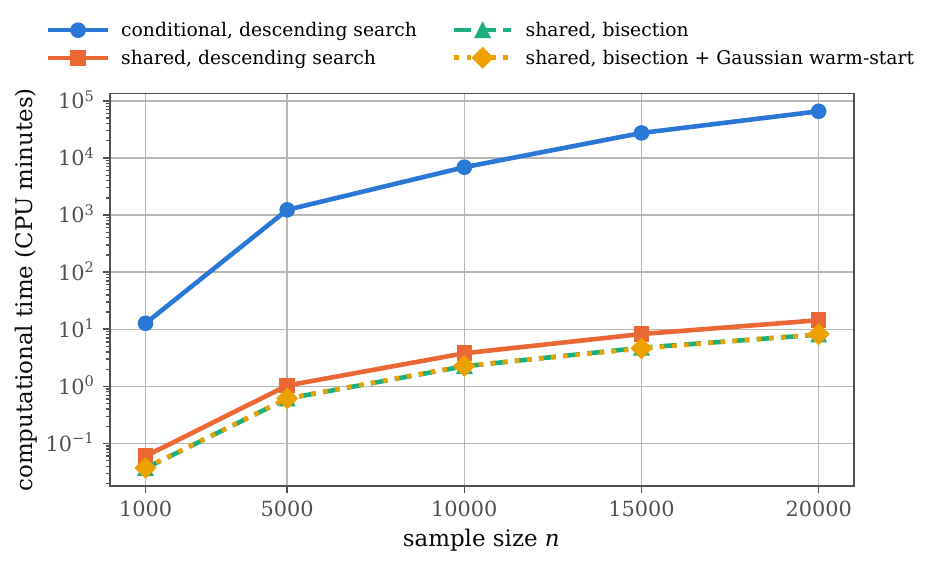}
\caption{Two dimensions.}
\end{subfigure}

\caption{CPU time in minutes, on a logarithmic scale, for
the four search methods in Section~\ref{subsec:exact_global_test}.
For methods using shared randomization draws, we report
the median time over $10$ runs. Conditional
search is slower, so we report its time
on only one run per sample size.}
\label{fig:runtime-efficiency}
\end{figure}

\clearpage
\endgroup

\clearpage

\subsubsection{Computational time}
\label{sec:exp-runtime}

We compare the computational time of four search methods
from Table~\ref{tab:computational_costs}: conditional
descending search, shared descending search, shared
bisection, and shared bisection with a Gaussian warm-start.
The one-dimensional design is the same as above.
For the two-dimensional timings, the zero-effect region
is a disc of area $0.1$ centred at $(0.5,0.5)$, containing
about $10\%$ of the units. We use the centred statistic
in both dimensions.
We measure CPU time on AMD EPYC~7J13 processors in one
dimension and Intel Xeon Platinum~8358 processors in two
dimensions.

Figure~\ref{fig:runtime-efficiency} shows that shared
descending search is up to several thousand times faster
than conditional descending search in both dimensions. 
Bisection further reduces computation substantially in one
dimension, with a smaller gain in two dimensions. In one
dimension, cumulative sums along a single ordering allow
each selected length to be tested cheaply. In two dimensions,
the implementation first computes cumulative sums for every
centre and neighbourhood size, so bisection saves
less computation.
The Gaussian warm-start saves little at $M=500$, but gives
larger savings when both the sample size $n$ and the number
of draws $M$ are large.

\subsection{Real-data experiments}

\subsubsection{The Gerber--Green--Larimer social-pressure experiment}
\label{sec:exp-ggl-main}

We analyse the Gerber--Green--Larimer voter-turnout
experiment \citep{gerber2008social}, comparing households
assigned to receive no mailing (Control) with those
assigned to receive a social-pressure mailing (Neighbors).
The mailing listed whether members of the household and
their neighbours had voted in previous elections and
announced that updated records would be circulated after
the election. It aimed to encourage voting by making
households aware that their neighbours could see whether
they voted.
Our sample contains $n=119{,}999$ households from
$10{,}000$ groups within which treatment was randomized.
The outcome is the household turnout rate in the 2006
primary election. We order households by their mean age,
breaking ties by group and household identifiers.

We implement our method with the centred statistic, $\alpha=0.05$ and
$M=5000$ shared permutations. Each permutation preserves
the observed Control and Neighbors counts within every
group, following the recorded assignment design.

\begin{figure}[t]
\centering
\includegraphics[width=0.66\textwidth]{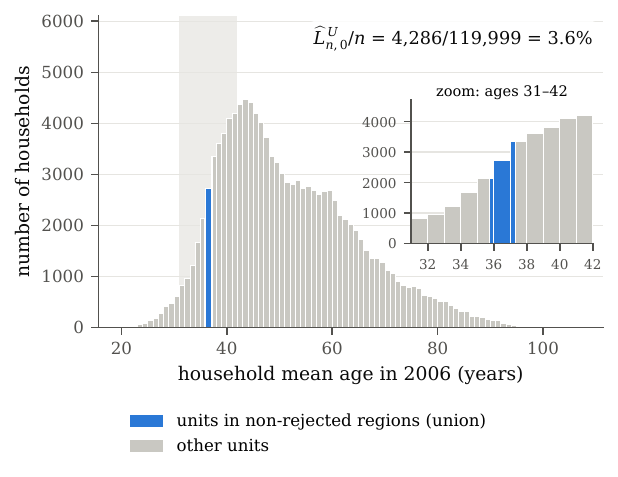}
\caption{Results from the Gerber--Green--Larimer experiment.
The histogram shows the distribution of the average age
of household members in 2006 across the $119{,}999$
households analysed.
The $95\%$ upper confidence bound for the largest
zero-effect block is $\widehat L_{n,0}^{\,U}=4286$
households ($3.6\%$ of the sample).
Blue marks the union of non-rejected blocks of this size.
The inset enlarges ages $31$--$42$, shaded in the main panel.}
\label{fig:real-ggl}
\end{figure}

The global sharp null is rejected, and the mean household
turnout difference is $8.48$ percentage points
(standard error: $0.34$ percentage points).
Figure~\ref{fig:real-ggl} shows the union of non-rejected
blocks at the largest non-rejected integer length,
$\widehat L_{n,0}^{\,U}=4286$ households
($3.6\%$ of the sample).
This $95\%$ upper confidence bound for $L_0$ complements
the global rejection by ruling out contiguous blocks of
unaffected households containing more than $3.6\%$ of
the sample along the age ordering.
We find only one non-rejected block of size
$\widehat L_{n,0}^{\,U}$, containing households with
mean ages between $35.5$ and $37$ years. Taken together,
these findings suggest that the mailing increases turnout
on average and that its effects extend broadly across
household ages.

\subsubsection{ACTG~175 HIV clinical trial}
\label{sec:exp-actg-main}
We analyse data from $1054$ patients in the ACTG~175
HIV clinical trial \citep{hammer1996}, comparing zidovudine alone (control) with the combination
of zidovudine and didanosine (treatment).
We order the sample by baseline CD4 count and consider
the 20-week CD4 count and its change
from baseline as outcomes.

We implement our method with the centred statistic, $\alpha=0.05$ and $M=5000$ shared permutations.
We permute treatment assignments within three groups
defined by patients' antiretroviral treatment histories,
preserving each group's observed treatment and control
counts and assuming that all such assignments are equally likely.

\clearpage
\begin{figure}[!p]
\centering
\captionsetup{font=small,skip=4pt}

\includegraphics[
  width=\textwidth,
  height=0.35\textheight,
  keepaspectratio
]{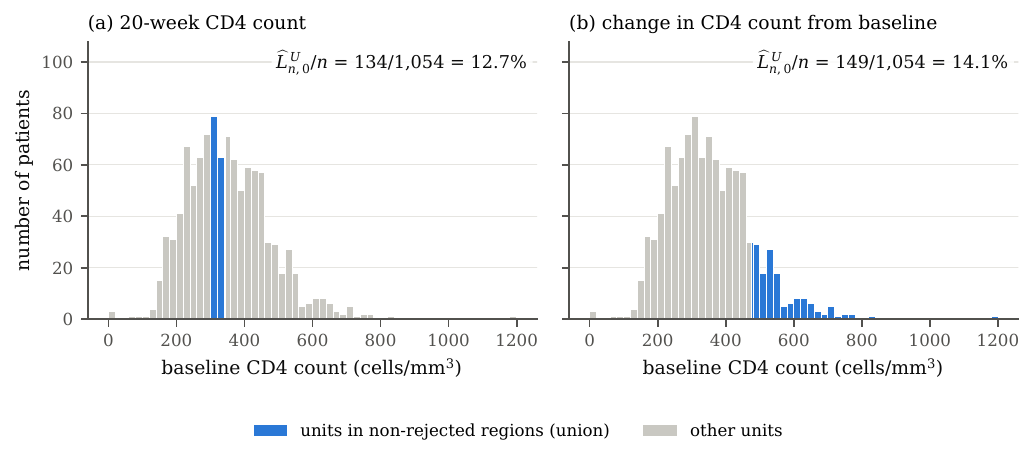}
\caption{Results from ACTG~175 with patients ordered by
baseline CD4 count. The histogram shows the empirical distribution
of baseline CD4 counts. The $95\%$ upper confidence bounds
for the largest zero-effect block are
(a) $\widehat L_{n,0}^{\,U}=134$ patients ($12.7\%$)
for the 20-week CD4 count and
(b) $149$ patients ($14.1\%$) for its change from baseline.
Blue marks the non-rejected block of the reported size
$\widehat L_{n,0}^{\,U}$ in each panel. Bins with at least
three quarters of their patients in the block are drawn
entirely blue.}
\label{fig:real-actg}

\vspace{1.5em}

\includegraphics[
  width=\textwidth,
  height=0.35\textheight,
  keepaspectratio
]{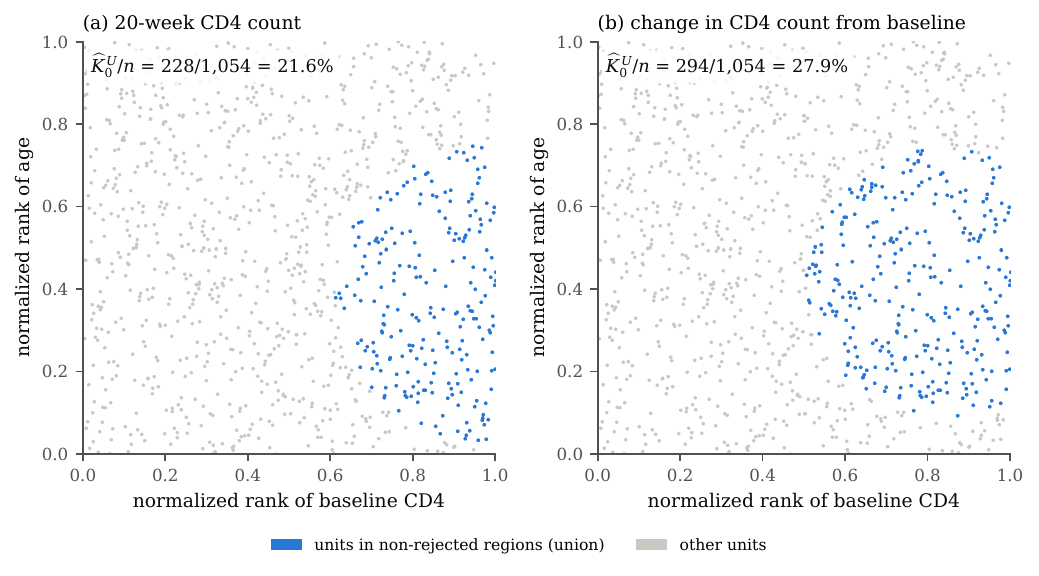}
\caption{Results from ACTG~175 with baseline CD4 count
and age as covariates.
Points are the normalized ranks of these two covariates.
The $95\%$ upper confidence bounds for the largest
zero-effect neighbourhood are
(a) $\widehat K_0^{\,U}=228$ patients ($21.6\%$)
for the 20-week CD4 count and
(b) $294$ patients ($27.9\%$) for its change from baseline.
Blue marks the union of non-rejected neighbourhoods
of the size $\widehat K_0^{\,U}$ in each panel.}
\label{fig:real-actg-2d}

\end{figure}
\clearpage

The $95\%$ upper confidence bound for $L_0$ is
$\widehat L_{n,0}^{\,U}=134$ patients
($12.7\%$ of the sample) for the 20-week CD4 count
and $149$ patients ($14.1\%$) for the change from baseline.
These bounds rule out large contiguous groups of
unaffected patients over the baseline CD4.
For each outcome, Figure~\ref{fig:real-actg} shows in blue
the only non-rejected block at the reported size.
The Baseline CD4 counts in this block range from $302$ to
$341$ cells/mm$^3$ for the 20-week outcome and from
$474$ to $1199$ cells/mm$^3$ for the change from baseline.

We define nearest-neighbour regions using baseline
CD4 count and age (Figure~\ref{fig:real-actg-2d}).
Using the same randomization draws and searching over all
integer sizes, we obtain $95\%$ upper confidence bounds
of $\widehat K_0^{\,U}=228$ patients ($21.6\%$) for the
20-week CD4 count and $294$ patients ($27.9\%$) for the
change from baseline.
Adding age changes the candidate regions and the target:
these bounds concern the largest zero-effect neighbourhood
defined jointly by baseline CD4 count and age.

\section{Discussion}
\label{sec:discussion}

We develop randomization inference for the reach of
treatment effects, providing a finite-sample valid upper
confidence bound for the size of the largest zero-effect
region. The bound and displays of non-rejected regions
show how broadly treatment effects extend across covariate
values, complementing existing randomization inference
methods for average effects and distributions of individual effects.

Obtaining this upper confidence bound requires running  randomization
tests across many overlapping regions at all
locations and region sizes.
We make this construction computationally efficient by
sharing randomization draws and reusing cumulative sums
for linear statistics. Our theoretical analysis shows how the number of affected units, the sizes of their effects, and the geometry of the
zero-effect set affect the tightness of the bound.
It establishes a minimax optimal rate, up to a logarithmic
factor, for the bound's excess above the true size.  Applications to experimental data from several domains
demonstrate how the method moves beyond detecting a
treatment effect to assessing how broadly it extends
across covariate values.

Several extensions could broaden the scope of our method.
Studying regions where all effects satisfy
$|\tau_i|\le\delta_0$ for a specified tolerance would allow
researchers to bound the size of the largest region with
no practically meaningful treatment effect. This requires valid tests of composite local nulls.
Moreover, detecting mixtures of positive and negative
effects without cancellation calls for local statistics
beyond the linear statistic considered here. These statistics
may not allow shared calculations,
creating new computational challenges.
Finally, extending inference from region
size to region location would require joint error control across
candidate regions to provide confidence guarantees for
where zero-effect regions may lie.

\bibliographystyle{plainnat}
\bibliography{references}

\clearpage

\appendix
\appendixpage

\section{Outcome adjustment in randomization tests}
\label{app:conditional_adjustment}

Conditional tests and shared global randomization tests differ
in which fitted outcome adjustments they can use. 
Throughout this appendix, the tested block $B$ is chosen using
pre-treatment information, and the fitting rule does not use
treatment assignments. Any randomness in fitting is held fixed.

\subsection{Why the local null matters}
\label{app:resid-outcomefit}
\label{app:resid-crossfit}

Consider the alternative adjusted statistic
\[
T_B^{\mathrm{adj}}(z_B;Y_{[n]})
:=\sum_{i\in B}(z_i-\pi)(Y_i-\widehat Y_i),
\]
where $\widehat Y_i$ is a prediction from a model
fitted to all observed covariates and outcomes. The observed statistic uses
$z_B=Z_B$; the randomized statistics hold the observed outcomes
and fitted predictions fixed and vary $z_B$.

Under the global Fisher sharp null, every observed outcome is
unchanged under reassignment. The fitted predictions therefore
remain fixed as well, so this adjustment preserves the usual
randomization argument without requiring a correctly specified
regression model.
Under the local sharp null $H_{0,B}$, only the outcomes within
$B$ are fixed. Outside outcomes may change with the assignment
and affect the fitted predictions. Thus, a fit can depend on
the assignment even when treatment is not included as a predictor.
Freezing that fit while drawing new global assignments does not
necessarily make the observed and randomized statistics exchangeable.

\subsection{Why conditional tests permit the adjustment}
\label{app:resid-condblock}

Let $Q_D$ denote the known treatment assignment distribution,
and let $Z_{-B}$ denote the assignments outside $B$.
Conditional on $Z_{-B}$ and under $H_{0,B}$,
outcomes inside $B$ are fixed by the null, and outcomes
outside $B$ are fixed by their assignments.
All observed outcomes and the fitted predictions are
therefore fixed.

Draw $Z_B^{(1)},\dots,Z_B^{(M)}$ independently
from $Q_D(Z_B\mid Z_{-B})$, independently of the observed
$Z_B$ conditional on $X_{[n]},Y_{[n]}(0),Y_{[n]}(1)$ and $Z_{-B}$. Under complete randomization, fixing $Z_{-B}$ fixes the number
of treated units in $B$, $K_B=n_1-\sum_{i\notin B}Z_i$. The conditional distribution is
uniform over the $\binom{|B|}{K_B}$ assignments that treat
exactly $K_B$ units in $B$. This is the usual within-block
conditional permutation test.
Define 
\[
p_{B,M}^{\mathrm{cond}}
:=\frac{1+\sum_{m=1}^M
\one\{T_B^{\mathrm{adj}}(Z_B^{(m)};Y_{[n]})
\ge T_B^{\mathrm{adj}}(Z_B;Y_{[n]})\}}{M+1}.
\]

\begin{proposition}
\label{prop:conditional_fitted_adjustment}
Suppose $H_{0,B}$ holds. The $p$-value
$p_{B,M}^{\mathrm{cond}}$ defined above satisfies
\[
\P\!\left\{p_{B,M}^{\mathrm{cond}}\le\alpha
\,\middle|\,X_{[n]},Y_{[n]}(0),Y_{[n]}(1),Z_{-B}\right\}\le\alpha,
\qquad \alpha\in[0,1].
\]
\end{proposition}

The proof is given in Appendix~\ref{app:proof-validity}.
Averaging over $Z_{-B}$ gives
\[
\P\!\left\{p_{B,M}^{\mathrm{cond}}\le\alpha
\,\middle|\,X_{[n]},Y_{[n]}(0),Y_{[n]}(1)\right\}\le\alpha.
\]
These valid local $p$-values can therefore be combined by the
same maximum-$p$ construction and inversion used in the main manuscript.

\subsection{A counterexample for global randomization}
\label{app:pooled_mean_counterexample}

Consider six units, exactly three of which are treated. Let
$B=\{1,2,3\}$ and set
\[
Y_i(0)=Y_i(1)=0\quad(i\in B),\qquad
Y_i(0)=-1,\quad Y_i(1)=1\quad(i\notin B).
\]
The local null on $B$ is true. Fit an intercept to all six
observed outcomes, so $\widehat Y_i=\bar Y$ for every unit,
where $\bar Y=n^{-1}\sum_{i=1}^nY_i$. The adjusted statistic is
\[
T_B^{\mathrm{adj}}(Z_B;Y_{[n]})
=\sum_{i\in B}(Z_i-\tfrac12)(Y_i-\bar Y).
\]
Writing $K=\sum_{i\in B}Z_i$, we have
\[
\bar Y=\frac{3-2K}{6},\qquad
T_B^{\mathrm{adj}}(Z_B;Y_{[n]})=\frac{(K-3/2)^2}{3}.
\]
For a new global assignment $Z'$, let $K'=\sum_{i\in B}Z_i'$.
With the fitted mean held fixed,
\[
T_B^{\mathrm{adj}}(Z_B';Y_{[n]})
=\frac{(K'-3/2)(K-3/2)}{3}.
\]
The count $K'$ equals $0,1,2,3$ with probabilities
$1/20,9/20,9/20,1/20$, respectively. Holding the observed
count $K$ fixed, the exact randomization $p$-value is
\[
p_B^{\mathrm{global}}
:=\P_{K'}\left\{
(K'-3/2)(K-3/2)\ge(K-3/2)^2
\right\}
=\begin{cases}
1/20,&K\in\{0,3\},\\
1/2,&K\in\{1,2\}.
\end{cases}
\]
Thus,
\(
\P\{p_B^{\mathrm{global}}\le0.05\}
=0.10
\)
under $H_{0,B}$.
In contrast, centering by $\bar Y_B$ gives a statistic of zero
and a $p$-value of one in this example.

\section{Warm-start using a Gaussian approximation}
\label{app:gaussian_warmstart}
\label{subsec:bisection_mc}
\label{par:warmstart}

Under the monotonicity in \eqref{eq:mc_monotonicity}
we use a Gaussian approximation to find a tentative bracket
before checking it with shared permutations. Under complete
randomization, with the within-block outcomes held fixed, the
statistic $T_B^{(m)}$ under a simulated assignment has mean zero
and variance
\[
\sigma_B^2
:=\var(T_B^{(m)})
=\frac{n\pi(1-\pi)}{n-1}
\sum_{i\in B}(Y_i-\bar Y_B)^2.
\]
The formula simplifies because
$\sum_{i\in B}(Y_i-\bar Y_B)=0$. Approximating this randomization
distribution by $N(0,\sigma_B^2)$ gives the one-sided Gaussian
$p$-value
\[
\widehat p_r
:=\max_{B\in\mathcal B_r}
\left\{1-\Phi\!\left(\frac{T_B}{\sigma_B}\right)\right\},
\]
where $\Phi$ is the standard normal distribution function and
$T_B$ is the observed statistic. If $\sigma_B^2=0$, then every
within-block outcome is equal and both $T_B$ and $T_B^{(m)}$
are zero; we set the corresponding Gaussian block $p$-value to $1$.
Using
\[
\sum_{i\in B}(Y_i-\bar Y_B)^2
=\sum_{i\in B}Y_i^2
-\frac{1}{|B|}\left(\sum_{i\in B}Y_i\right)^2,
\]
we compute the variance from cumulative sums of $Y_i$ and $Y_i^2$.
Together with the cumulative sums for $T_B$, this gives
an $O(n)$ cost for evaluating $\widehat p_r$ at each length.

\paragraph{Finding a tentative bracket.}
The Gaussian search considers all lengths $r\in[n]$ and
starts at $r=1$. If its $p$-value exceeds $\alpha$, we
double the length, capping it at $n$. We continue until
the Gaussian $p$-value is at most $\alpha$ or we reach $n$.

If the first rejection occurs after the initial length,
we let $r_+$ be that length and $r_-$ the length evaluated
immediately before it. These endpoints satisfy
\[
r_-<r_+,\qquad
\widehat p_{r_-}>\alpha,\qquad
\widehat p_{r_+}\le\alpha.
\]
We narrow the bracket by bisection. At each step,
we evaluate the Gaussian $p$-value at
$\lfloor(r_-+r_+)/2\rfloor$. This length replaces $r_-$
if its $p$-value exceeds $\alpha$, and $r_+$ otherwise.
We stop when $r_+=r_-+1$.
If length $1$ is rejected by the
Gaussian approximation, or no Gaussian rejection is found,
we use the direct Monte Carlo search described below.

We evaluate the Monte Carlo $p$-values at the two endpoints.
If they satisfy
\begin{equation}
\tilde p_{r_-}>\alpha,
\qquad
\tilde p_{r_+}\le\alpha,
\label{eq:mc_bracket_conditions}
\end{equation}
we report $r_-$ as the upper confidence bound for $L_0$.
The endpoints are adjacent, and the monotonicity in
\eqref{eq:mc_monotonicity} ensures that every larger length
is rejected. 

If either inequality fails, we use direct Monte Carlo
bisection.
We first check length $1$. If it is rejected, monotonicity
implies that all lengths are rejected, and we report zero.
Otherwise, we check length $n$. If it is not rejected, we
report $n$. In the remaining case, lengths $1$ and $n$
form a valid Monte Carlo bracket. We use bisection
to narrow this bracket, replacing the lower endpoint
after a non-rejection and the upper endpoint after a
rejection. Once the endpoints are adjacent, we report
the lower endpoint. The same direct search is used when
the Gaussian search does not produce a bracket.

\paragraph{Computational cost.}
Let $q$ be the number of Gaussian evaluations and $k$ the
total number of Monte Carlo evaluations, including any
fallback search. The total cost is
$C_{\mathrm{warm}}=O(nq+Mnk).$
Here $q=O(\log n)$. If the adjacent Gaussian
endpoints satisfy \eqref{eq:mc_bracket_conditions},
then $k=2$ and $C_{\mathrm{warm}}=O(n\log n+Mn).$
A poor approximation may require the fallback search,
for which $k=O(\log n)$. The cost then becomes
$O(n\log n+Mn\log n)$, removing the saving in Monte Carlo
evaluations relative to direct bisection. The Gaussian approximation affects the search cost.
The validity of the reported bound follows from the
Monte Carlo tests and the monotonicity assumption:
under \eqref{eq:mc_monotonicity}, the search returns
the same bound as a search over all lengths $r\in[n]$.

\section{Dimension and neighbourhood geometry}
\label{app:knn_dimension}

For each of the $n$ observed centres, we form one neighbourhood
at each size $k=1,\dots,n$. There are therefore at most $n^2$
distinct candidate neighbourhoods in any dimension. 
Dimension can nevertheless affect how many affected units a growing neighbourhood contains, and hence
the exponent $\nu$ in \eqref{eq:knn_geometric_count}, which controls the lower
bound on the number of units sufficiently far from $\mathcal Z_0$.

\begin{figure}[t]
\centering
\includegraphics[width=0.9\linewidth]{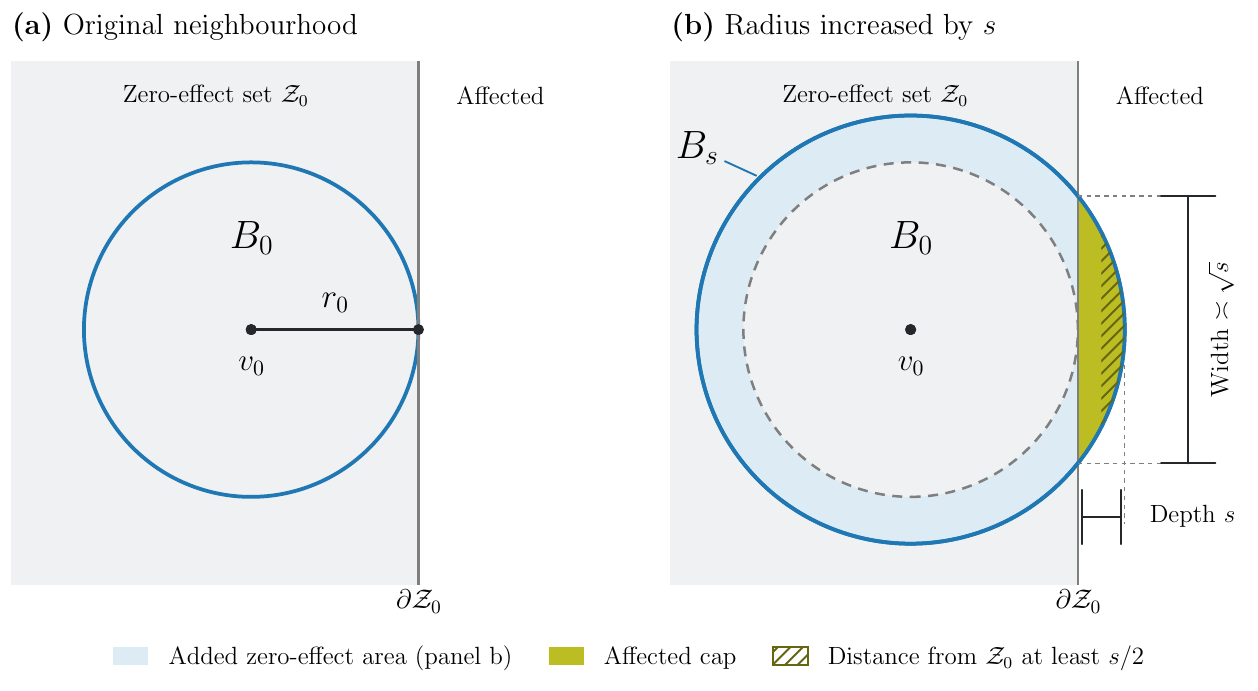}
\caption{Expansion across a zero-effect boundary in two
dimensions. (a) The ball $B_0$ lies in $\mathcal Z_0$ and
touches its boundary. (b) Increasing the radius from $r_0$ to $r_0+s$ adds
zero-effect area (blue) and an affected cap (olive).
The hatched part lies at least $s/2$ from $\mathcal Z_0$.
For fixed $r_0$ and small $s$, the band $B_s\setminus B_0$
has area of order $s$. Its affected part, the cap
$B_s\setminus\mathcal Z_0$, has area of order $s^{3/2}$.}
\label{fig:knn_boundary_geometry}
\end{figure}

\Cref{fig:knn_boundary_geometry} illustrates this connection. Let $P_V$ be a continuous
covariate distribution in rank space. Suppose all points to the
left of a vertical boundary $\partial\mathcal Z_0$ have zero
treatment effect, while points to the right are affected. Consider a ball $B_0$
of radius $r_0>0$ entirely inside $\mathcal Z_0$ and touching
this boundary at one point in
\Cref{fig:knn_boundary_geometry}(a).
Let $B_s$ have the same centre and radius $r_0+s$.
Assume these balls lie inside the covariate support, where
the density is bounded above and below by positive constants.

In \Cref{fig:knn_boundary_geometry}(b), the cap has horizontal
depth $s$. Its width is the distance between the two points
where the enlarged circle meets the boundary. Since the
centre is at distance $r_0$ from the boundary, the
Pythagorean theorem gives
\[
\text{width}
=2\sqrt{(r_0+s)^2-r_0^2}
=2\sqrt{2r_0s+s^2}
\asymp\sqrt{s},
\]
for fixed $r_0>0$ as $s\downarrow0$.
In $d\ge2$ dimensions, the cross-section
$B_s\cap\partial\mathcal Z_0$ is a $(d-1)$-dimensional
ball with diameter of order $\sqrt{s}$. Its
$(d-1)$-dimensional volume is therefore of order
$s^{(d-1)/2}$.
The cap $B_s\setminus\mathcal Z_0$ extends a distance $s$
beyond the boundary. Integrating its cross-sectional
volumes over this depth gives
\[
\operatorname{Vol}_d(B_s\setminus\mathcal Z_0)
\asymp s^{(d+1)/2}.
\]
The full added band $B_s\setminus B_0$ has volume proportional
to $(r_0+s)^d-r_0^d\asymp s$ for fixed $r_0>0$ as
$s\downarrow0$. Thus, under the density bounds above,
$u_s:=P_V(B_s\setminus B_0)\asymp s$.
As indicated by the hatched area, a fixed fraction of the
cap's volume lies at distance at least $s/2$ from
$\mathcal Z_0$. Consequently, as $s\downarrow0$,
\begin{equation}
P_V\{v\in B_s:
\operatorname{dist}(v,\mathcal Z_0)\ge s/2\}
\asymp s^{(d+1)/2}
\asymp u_s^{(d+1)/2}.
\label{eq:knn_cap_mass}
\end{equation}
Since $s/2\asymp u_s$, the enlarged ball contains affected
mass of order $u_s^{(d+1)/2}$ at distance of order $u_s$
from $\mathcal Z_0$. The affected mass grows as the $(d+1)/2$ power of the
added mass. This corresponds to $\nu=(d+1)/2$ in
\eqref{eq:knn_geometric_count}, provided the same lower
bound holds uniformly over the sample neighbourhoods.

Applying this calculation to $K_0$ requires the count condition
to hold for every sample neighbourhood larger than $K_0$.
If \eqref{eq:knn_geometric_count} holds uniformly with
$\nu=(d+1)/2$, then $\gamma=\frac{d+1}{2}+\beta$, defined above \eqref{eq:knn_signal_growth}.
For fixed $\zeta$ and sufficiently large $M$ and $n$,
\Cref{thm:knn_power} gives, with probability at least
$1-\zeta$ conditional on $X_{[n]}$,
\[
\frac{\widehat K_0^{\,U}-K_0}{n}
\lesssim
\left(\frac{\log n}{n}\right)^{1/(d+1+2\beta)}.
\]
This illustrates one way that dimension can slow the
decrease of the excess-size bound. Other shapes of
$\mathcal Z_0$ or of the covariate support can lead to
different values of $\nu$.

\paragraph{Examples of effect patterns.}
The class $\mathcal F_n(\lambda,\gamma)$ defined before
Theorem~\ref{thm:knn_minimax_lower} allows different
zero-effect sets and decompositions $\gamma=\nu+\beta$.
This class includes spatial models with abrupt or gradual
changes in treatment effects.
For example, let $\mathcal Z_0$ be a ball in rank space
centred at an observed point $v_{i_0}$ and containing
$m\in\{1,\dots,n-1\}$ units. Consider piecewise constant
treatment effects 
\[
\tau_j=\tau_*\one\{v_j\notin\mathcal Z_0\},
\qquad 0<\tau_*\le\tau_{\max}.
\]
The $m$ unaffected units form $N_{i_0}^{(m)}$, so $K_0=m$.
Every neighbourhood of size $k>m$ contains at least
$k-m$ affected units, giving
\[
\bar\tau_{i,k}
\ge\tau_*\frac{k-K_0}{n}
\ge\tau_*\left(\frac{k-K_0}{n}\right)^\gamma.
\]
Taking $\beta=0$, $\nu=\gamma$, $a=1$,
$\lambda_\tau=\tau_*$, and $c=\lambda/\tau_*$ verifies
membership in $\mathcal F_n(\lambda,\gamma)$ whenever
$\lambda\le\tau_*$.
When the covariate sample satisfies
\eqref{eq:knn_geometric_count}, the class also includes
continuous treatment effects that increase with distance
from a zero-effect region:
\[
\tau(v)=\tau_*\operatorname{dist}(v,\mathcal Z_0)^\beta,
\qquad \beta>0,
\]
with $\tau_*>0$ chosen so that the  effects are
bounded by $\tau_{\max}$.
These are two examples of 
$\tau_{[n]}\in\mathcal F_n(\lambda,\gamma)$. The class also includes
other patterns satisfying Assumption~\ref{ass:knn_geometry}.

\section{Convergence in three settings}
\label{app:exp-rates}

We study how treatment-effect strength and the shape of
the zero-effect region affect convergence. We simulate data
using the outcome model and complete randomization in
Section~\ref{sec:exp-overview}. The randomization tests use
the centred statistic in \eqref{eq:centered_stat}, with
$\alpha=0.05$ and $M=500$ shared draws. We search over
all integer sizes and run $100$ repetitions for each
$n\in\{500,1000,2000,4000,8000,16000\}$.
Let $g_{i\ell}$ be the rank of unit $i$ in coordinate $\ell$
and write $v_i=(g_{i1}/n,\dots,g_{id}/n)$.
We consider the following three effect patterns: the first
two in both one and two dimensions.

{\it Constant effects outside a central region.}
Let $m=\lfloor0.10n\rfloor$ and
$q=\lfloor(n-m)/2\rfloor$. Define the treatment effect of
unit $i$ by
\[
\tau_i=\one\{i\notin I_0\}.
\]
In one dimension, $I_0=\{i:q<g_{i1}\le q+m\}$ is the
central block of $m$ units. In two dimensions,
$I_0=N_{i_0}^{(m)}$ contains the $m$ units nearest to
unit $i_0$, including $i_0$ itself, using Euclidean distance
between normalized covariate ranks. The centre $i_0$ is
chosen as in Section~\ref{sec:exp-overview}.
Thus, effects are zero inside $I_0$ and one outside it.

{\it Effects increasing from the boundary.}
Use the same region $I_0$, but let effects grow linearly
with distance to its nearest unit. Specifically, define
\[
\tau_i=\frac{\rho_i}{\rho_{\max}},      \qquad
\rho_i:=\min_{j\in I_0}\|v_i-v_j\|_2,   
\qquad
\rho_{\max}:=\max_{i\in[n]}\rho_i
.
\]
Effects are zero on $I_0$ and reach one at the largest
distance. In both central-region patterns, exactly $m$
units are unaffected, so $K_0=m$, with $K_0=L_0$ in one dimension.

\begin{figure}[t]
\centering
\includegraphics[width=\textwidth]{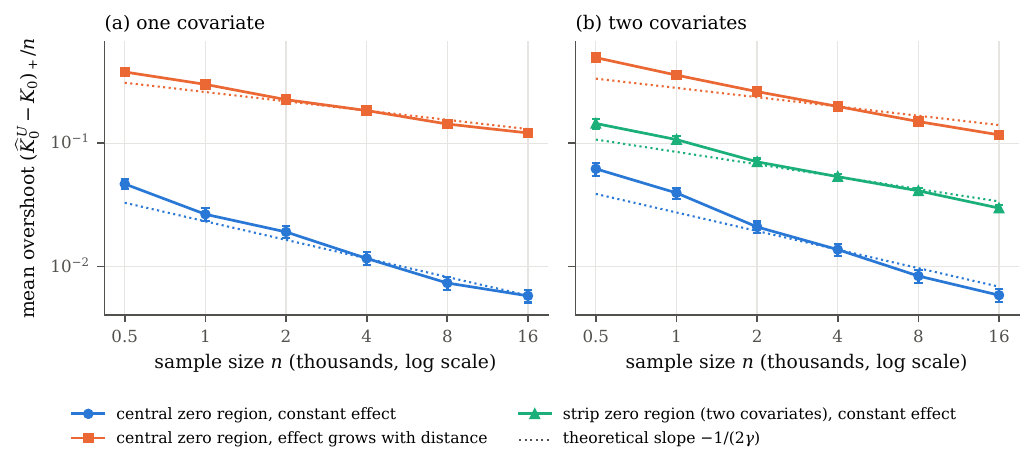}
\caption{Convergence under different treatment-effect
patterns. Mean positive excess
$(\widehat K_0^{\,U}-K_0)_+/n$, with $\pm2$ standard errors
over $100$ repetitions, on logarithmic axes.
Dotted lines show reference slopes $-1/2$ for constant
effects outside the central region, $-1/4$ for effects
that grow linearly with distance, and $-1/3$ for the strip.}
\label{fig:rates}
\end{figure}

{\it A zero-effect band.}
In two dimensions, the unaffected units are those with
the lowest $20\%$ of first-coordinate ranks, regardless
of their second-coordinate ranks. Define the treatment effect of unit $i$ as $\tau_i=\one\{i\notin I_{\mathrm{band}}\}$, where
\[
 I_{\mathrm{band}}
=\{i:g_{i1}\le\lfloor0.20n\rfloor\},
\]
We compute the size of the largest zero-effect neighbourhood as
\[
K_0=\max\{k\in[n]:
\exists i\in[n],\ N_i^{(k)}\subseteq I_{\mathrm{band}}\}.
\]
The largest such neighbourhood need not contain the entire
zero-effect band.

\paragraph{Results.}
For each repetition, we report the positive excess
\[
\frac{(\widehat K_0^{\,U}-K_0)_+}{n},
\qquad x_+:=\max\{x,0\},
\]
where $\widehat K_0^{\,U}=\widehat L_{n,0}^{\,U}$ in one
dimension. Taking the positive part prevents underestimation
in some repetitions from offsetting overestimation in others.

Figure~\ref{fig:rates} shows that the mean excess decreases
with sample size in all five settings. It decreases fastest
for constant effects outside the central region and slowest
when effects grow gradually from the boundary.
For comparison with Theorem~\ref{thm:knn_power}, the dotted
lines have slopes $-1/(2\gamma)$, using $\gamma=1$ for
constant effects, $\gamma=2$ for gradually increasing
effects, and $\gamma=3/2$ for the band. The band value
is motivated by the geometric calculation in
Appendix~\ref{app:knn_dimension}. These are illustrative
comparisons: the theorem bounds the excess with high
probability, whereas the figure reports its mean.

\section{Technical proofs}
\label{app:theoretical_proofs}

\subsection{Validity}
\label{app:proof-validity}
\label{app:proof-length-bound}
\label{app:proof-knn-validity}
\label{app:proof-conditional-adjustment}

This subsection proves Proposition~\ref{prop:length_bound_validity},
Theorem~\ref{thm:knn_validity}, and
Proposition~\ref{prop:conditional_fitted_adjustment}.
Throughout, condition on $X_{[n]},Y_{[n]}(0),Y_{[n]}(1)$ and the
known design. Candidate regions and their ordering are therefore fixed.
All probabilities below use this conditioning; additional conditioning
is displayed when needed.  Fix $M\ge1$ and $\alpha\in[0,1]$.

\begin{proof}
\textbf{1. Local tests.}
Fix a region $B$ whose local sharp null holds. Then
\[
Y_j(z_j)=Y_j^*:=Y_j(0)=Y_j(1),\qquad j\in B,
\]
so the centered statistic is the fixed function
\[
t_B(z)=\sum_{j\in B}(z_j-\pi)(Y_j^*-\bar Y_B^*),
\qquad \bar Y_B^*=\frac1{|B|}\sum_{j\in B}Y_j^*.
\]
Let $Z^{(0)}=Z$ and draw
$Z^{(1)},\ldots,Z^{(M)}$ independently
from the same distribution as $Z$. The vector with entries
$V_m:=t_B(Z^{(m)})$, $m=0,\ldots,M$, is exchangeable. 

Let $R_m$ be the rank of $V_m$ among $V_0,\ldots,V_M$,
ordered from largest to smallest, with tied values assigned
the largest rank occupied by their group. Then
\[
R_m=\sum_{\ell=0}^M\one\{V_\ell\ge V_m\},
\qquad \tilde p_{B,M}=\frac{R_0}{M+1}.
\]
A value in position $j$, counting from largest to smallest,
has rank at least $j$, because the first $j$ values are all
at least as large as it. Therefore, only the first $a$
positions can have rank at most $a$, so
\[
\sum_{m=0}^M\one\{R_m\le a\}\le a,
\qquad a\in\{0,\ldots,M+1\}.
\]
Exchangeability and $a=\lfloor\alpha(M+1)\rfloor$ give
\begin{equation}
\begin{aligned}
\P\{\tilde p_{B,M}\le\alpha\}
&=\frac1{M+1}\sum_{m=0}^M\P\{R_m\le a\}\\
&=\frac1{M+1}\E\!\left[\sum_{m=0}^M
   \one\{R_m\le a\}\right]
\le\frac{a}{M+1}\le\alpha.
\end{aligned}
\label{eq:proof-local-validity}
\end{equation}
This includes ties and uses no null hypothesis outside $B$.
For exact randomization, list the distinct values of $t_B(Z)$ as
$v_1>\cdots>v_J$, with positive probabilities $q_1,\ldots,q_J$.
If $t_B(Z)=v_j$, its upper-tail $p$-value is
$p_B=\sum_{\ell\le j}q_\ell$. Thus
\[
\P\{p_B\le\alpha\}
=\sum_{j:\,\sum_{\ell\le j}q_\ell\le\alpha}q_j
\le\alpha.
\]
The inequality holds because the selected indices form an initial
segment. Hence both exact and Monte Carlo local tests are valid.

\textbf{2. Conditional tests with fitted adjustment.}
For Proposition~\ref{prop:conditional_fitted_adjustment}, condition
additionally on $Z_{-B}$. Under the local null,
\[
Y_j=
\begin{cases}
Y_j^*,&j\in B,\\
Y_j(Z_j),&j\notin B,
\end{cases}
\qquad\Longrightarrow\qquad
Y_{[n]}\text{ and }\widehat Y_{[n]}\text{ are fixed}.
\]
Here the fit uses only covariates and observed outcomes, with any
independent fitting randomness held fixed. Therefore
\[
z_B\longmapsto t_B^{\mathrm{adj}}(z_B)
:=\sum_{j\in B}(z_j-\pi)(Y_j-\widehat Y_j)
\]
is a fixed function. The observed $Z_B$ and simulated assignments
are independent draws from the distribution $Q_D(Z_B\mid Z_{-B})$.
The rank calculation in part~1 now gives
\[
\P\{p_{B,M}^{\mathrm{cond}}\le\alpha\mid Z_{-B}\}\le\alpha,
\qquad
\P\{p_{B,M}^{\mathrm{cond}}\le\alpha\}
=\E\!\left[\P\{p_{B,M}^{\mathrm{cond}}\le\alpha
  \mid Z_{-B}\}\right]\le\alpha.
\]
This proves the validity both conditional on $Z_{-B}$ and after averaging over $Z_{-B}$.
Under complete randomization, the conditional distribution $Q_D$ is given by 
\[
K_B=n_1-\sum_{j\notin B}Z_j,
\qquad
Q_D(z_B\mid Z_{-B})
=\binom{|B|}{K_B}^{-1}\one\!\left\{\sum_{j\in B}z_j=K_B\right\}.
\]

\textbf{3. Maximum $p$-values.}
Let $\mathscr R_k$ be any nonempty finite family of candidate regions
at size $k$ on a finite grid $\mathcal K$, and let $p_B$ be any
valid local $p$-value. Define
\[
\begin{aligned}
P_k&:=\max_{B\in\mathscr R_k}p_B,\\
k_0&:=\max\bigl(\{k\in\mathcal K:
  \exists B\in\mathscr R_k,\ H_{0,B}\text{ holds}\}\cup\{0\}\bigr),\\
\widehat k^{\,U}&:=\max\bigl(\{k\in\mathcal K:P_k>\alpha\}\cup\{0\}\bigr).
\end{aligned}
\]
If the union null at size $k$ is true, choose one true-null region
$B_0\in\mathscr R_k$. Then
\[
\{P_k\le\alpha\}\subseteq\{p_{B_0}\le\alpha\}
\quad\Longrightarrow\quad \P\{P_k\le\alpha\}\le\alpha.
\]
When $k_0>0$, apply this at the fixed true size $k_0$:
\begin{equation}
\{\widehat k^{\,U}<k_0\}
\subseteq\{P_{k_0}\le\alpha\}
\subseteq\{p_{B_0}\le\alpha\},
\qquad
\P\{\widehat k^{\,U}\ge k_0\}\ge1-\alpha.
\label{eq:proof-general-coverage}
\end{equation}
For $k_0=0$, coverage is automatic. An empty grid also gives both
quantities equal to zero. No independence across regions is needed.
Equivalently, since
\[
P_k^\uparrow:=\max_{s\in\mathcal K:\,s\ge k}P_s,
\qquad
P_k^\uparrow>\alpha
\ \Longleftrightarrow\ 
\exists s\in\mathcal K:\ s\ge k,\ P_s>\alpha.
\]
Taking the largest $k$ on either side returns $\widehat k^{\,U}$.
The choices
\[
\begin{array}{c|c|c}
\mathscr R_k&k_0&\widehat k^{\,U}\\ \hline
\mathcal B_k&L_0&\widehat L_0^{\,U}\\
\{N_i^{(k)}:i\in[n]\}&K_0&\widehat K_0^{\,U}
\end{array}
\qquad (\mathcal K=[n])
\]
prove Proposition~\ref{prop:length_bound_validity} and
Theorem~\ref{thm:knn_validity}. The same argument covers the rectangular construction
with $\mathscr R_k=\{B_R:|B_R|\ge k\}$ and
$k_0=L_{0,\mathrm{rect}}$.
\end{proof}

\subsection{Proof of Proposition~\ref{prop:knn_computational_cost}}
\label{app:nearest_neighbour_computation_proof}

\begin{proof}
Recall that for each centre $i$, let $j_i(1),\ldots,j_i(n)$ be the unit indices
ordered by Euclidean distance between the normalized rank vectors,
with the prescribed tie rule.
In the following, we write $Z^{(0)}=Z$ for the observed assignment.

\textbf{Neighbour orderings.}
Ranking the $n$ units in each of the $d$ coordinates takes
$O(dn\log n)$ time. Computing all pairwise squared distances
takes $O(dn^2)$ time, and sorting the $n$ distances at each
of the $n$ centres takes $O(n^2\log n)$ time. The total
preprocessing cost is therefore $O(dn^2+n^2\log n)$.
 We generate each simulated assignment by drawing a
uniform random permutation and treating the first $n_1$
units. This takes $O(n)$ time per assignment, or $O(Mn)$
for all $M$ assignments.

\textbf{Full search.}
For $m=0,\ldots,M$, maintain
\[
S_{i,k}:=\sum_{\ell=1}^kY_{j_i(\ell)},\qquad
C_{i,k}^{(m)}:=\sum_{\ell=1}^kZ_{j_i(\ell)}^{(m)},\qquad
U_{i,k}^{(m)}:=\sum_{\ell=1}^kZ_{j_i(\ell)}^{(m)}Y_{j_i(\ell)}.
\]
Starting from zero, the updates and statistic are
\[
\begin{aligned}
S_{i,k}&=S_{i,k-1}+Y_{j_i(k)},\\
C_{i,k}^{(m)}&=C_{i,k-1}^{(m)}+Z_{j_i(k)}^{(m)},\\
U_{i,k}^{(m)}&=U_{i,k-1}^{(m)}+Z_{j_i(k)}^{(m)}Y_{j_i(k)},\\
T_{i,k}^{(m)}&=U_{i,k}^{(m)}-\frac{S_{i,k}}{k}C_{i,k}^{(m)}.
\end{aligned}
\]
Thus all $M+1$ statistics and the count
$\sum_{m=1}^M\one\{T_{i,k}^{(m)}\ge T_{i,k}^{(0)}\}$
cost $O(M)$ per pair $(i,k)$. 
With $n^2$ pairs $(i,k)$, computing all observed and simulated
statistics and the corresponding Monte Carlo $p$-values costs
$n^2O(M)$. Generating the $M$ simulated assignments costs
$O(Mn)$. Taking the maximum $p$-value over the $n$ centres
at each of the $n$ sizes adds $nO(n)$. The total cost is therefore
\[
O(Mn)+n^2O(M)+nO(n)=O(Mn^2).
\]
\textbf{Descending search.} At $k=n$, all neighbourhoods equal $[n]$. Initialize their sums:
\[
S_{i,n}=\sum_jY_j,\qquad
C_{i,n}^{(m)}=\sum_jZ_j^{(m)},\qquad
U_{i,n}^{(m)}=\sum_jZ_j^{(m)}Y_j.
\]
This costs $O(Mn)$, including copying the values to all centres.

After this, each decrease in size uses
\[
\begin{aligned}
S_{i,k-1}&=S_{i,k}-Y_{j_i(k)},\\
C_{i,k-1}^{(m)}&=C_{i,k}^{(m)}-Z_{j_i(k)}^{(m)},\\
U_{i,k-1}^{(m)}&=U_{i,k}^{(m)}-Z_{j_i(k)}^{(m)}Y_{j_i(k)}.
\end{aligned}
\]
Updating the statistics, counts, and maximum costs $O(Mn)$
per size. For $t\ge1$ checked sizes, the total is
$O(Mn)+tO(Mn)=O(Mnt)$. The first non-rejected size is the largest
one because all larger sizes have already been rejected; if there
is none, the bound is zero. Adding the preprocessing cost proves both claims.
\end{proof}

\subsection{Proof of Theorem~\ref{thm:knn_power}}
\label{app:proof-knn-power}

\paragraph{Notation.}
For brevity, write $Y=Y_{[n]}$, $Z=Z_{[n]}$, and
$z=z_{[n]}$. For each candidate neighbourhood $B$, use
the shorthand
\[
T_B(z)
:=T_B\bigl(O_B(z_B)\bigr)
=\sum_{j\in B}(z_j-\pi)(Y_j-\bar Y_B),
\qquad
\bar Y_B:=\frac1{|B|}\sum_{j\in B}Y_j.
\]
The observed outcomes $Y_B$ are held fixed when evaluating
this function at different assignments. Thus,
$T_B(Z)=T_B(O_B)$ is the observed statistic, whereas
$T_B(Z')$ is the randomized statistic for an independent
assignment $Z'$ drawn from the same design.
We write $p_B$ and $\tilde p_B$ for the corresponding exact
and Monte Carlo $p$-values; when $B=N_i^{(k)}$, $\tilde p_B$ is $\tilde p_{i,k}$ from the main text.

\begin{proof}
We condition on $X_{[n]}$ throughout, so the neighbourhoods,
effects, and $K_0$ are fixed. Complete
randomization gives $\pi=n_1/n\in[\pi_{\min},1-\pi_{\min}]$.
Write
\[
\mu_j=\mu_0(X_j),\quad
s_Y=\sigma+\mu_{\max}+\tau_{\max},\quad
L=\log(32n^3/\zeta),\quad
\kappa=\pi_{\min}(1-\pi_{\min}).
\]
Let $\mathcal N$ be the family of distinct candidate
neighbourhoods, with $|\mathcal N|\le n^2$. We first control
the observed and randomized statistics, then the Monte Carlo
error, and finally the largest non-rejected size.
We assumed that $0\le\tau_j\le\tau_{\max}$ and
\[
A_{i,k}:=\sum_{j\in N_i^{(k)}}\tau_j
\ge n\lambda u_k^\gamma,
\qquad
u_k=\frac{k-K_0}{n}>0,
\quad \lambda=\lambda_\tau ca^\beta,
\quad \gamma=\nu+\beta\ge1.
\]
For $\beta>0$, this follows by multiplying the count
$cnu_k^\nu$ by the effect bound $\lambda_\tau(au_k)^\beta$.
For $\beta=0$, use the affected-unit count and the effect
bound $\lambda_\tau$. If $K_0=n$, the theorem is immediate.
Otherwise at least one effect is positive, so $s_Y>0$.

\paragraph{1. Simultaneous concentration.}
Fix $B\in\mathcal N$, write $k=|B|$, and let
$|a_j|\le A$. We derive a concentration bound using
Hoeffding's inequality and conditioning.

Let $\xi_1,\ldots,\xi_n$ be independent
$\operatorname{Bernoulli}(\pi)$ variables, where $\pi=n_1/n$.
Conditional on $\sum_{j=1}^n\xi_j=n_1$, their joint
distribution is exactly the complete-randomization
distribution of $Z_{[n]}$.
Each variable $(\xi_j-\pi)a_j$ has mean zero and lies
in an interval of length $|a_j|\le A$. Hence, for $A>0$,
Hoeffding's inequality gives
\[
\mathbb P\left\{
\left|\sum_{j\in B}(\xi_j-\pi)a_j\right|>t
\right\}
\le 2\exp\left(-\frac{2t^2}{kA^2}\right),
\qquad t>0.
\]
The count $\sum_{j=1}^n\xi_j$ has a
$\operatorname{Binomial}(n,\pi)$ distribution.
Then,
\[
\mathbb P\left\{\sum_{j=1}^n\xi_j=n_1\right\}
=\binom{n}{n_1}\pi^{n_1}(1-\pi)^{n-n_1}
\ge\frac1{n+1}.
\]
The inequality follows because $n_1$ is a most probable
count among the $n+1$ possible values. We use this simple
lower bound to simplify the subsequent bound.
Therefore,
\[
\begin{aligned}
\mathbb P\left\{
\left|\sum_{j\in B}(Z_j-\pi)a_j\right|>t
\right\}
&=
\mathbb P\left\{
\left|\sum_{j\in B}(\xi_j-\pi)a_j\right|>t
\,\middle|\,\sum_{j=1}^n\xi_j=n_1
\right\}\\
&\le
\frac{
\mathbb P\left\{
\left|\sum_{j\in B}(\xi_j-\pi)a_j\right|>t
\right\}
}{
\mathbb P\left\{\sum_{j=1}^n\xi_j=n_1\right\}
}\\
&\le 2(n+1)\exp\left(-\frac{2t^2}{kA^2}\right).
\end{aligned}
\]
Setting $t=A\sqrt{kx/2}$ yields
\begin{equation}
\mathbb P\left\{
\left|\sum_{j\in B}(Z_j-\pi)a_j\right|
>A\sqrt{\frac{kx}{2}}
\right\}
\le 2(n+1)e^{-x},
\qquad x>0.
\label{eq:knn_weighted_design_tail_proof}
\end{equation}
If $A=0$, the sum is identically zero and the bound
holds directly.

For a block $B$ with $k=|B|$, define its average baseline
mean by
\[
\bar\mu_B:=k^{-1}\sum_{j\in B}\mu_j,
\]
and its proportion of treated units by
\[
\bar Z_B:=k^{-1}\sum_{j\in B}Z_j.
\]
Also define
\[
A_B:=\sum_{j\in B}\tau_j,\qquad
D_B:=\sum_{j\in B}(Z_j-\pi),\qquad
E_B:=\sum_{j\in B}(Z_j-\pi)\tau_j.
\]
Here $A_B$ is the total treatment effect within $B$,
while $D_B$ and $E_B$ are the deviations of
$\sum_{j\in B}Z_j$ and $\sum_{j\in B}Z_j\tau_j$
from their respective expectations.

A neighbourhood is rejected when its observed statistic is
sufficiently large relative to its randomization distribution.
A large total treatment effect contributes positively to the
observed statistic, but random assignment can offset
this contribution. We next control these random deviations
simultaneously for all candidate neighbourhoods.

Apply \eqref{eq:knn_weighted_design_tail_proof} with
$x=L$ and $a_j=1$, $a_j=\tau_j$, and
$a_j=\mu_j-\bar\mu_B$, bounded respectively by
$1$, $\tau_{\max}$, and $2\mu_{\max}$ in absolute value.
For each choice and each neighbourhood, the resulting bound
fails with probability at most $2(n+1)e^{-L}$.
A union bound over the three choices and at most $n^2$
candidate neighbourhoods shows that the following event
has probability at least $1-6n^2(n+1)e^{-L}$:
\[
\mathcal E_1
:=
\bigcap_{B\in\mathcal N}
\left\{
\begin{aligned}
|D_B|&\le\sqrt{kL/2},\\
|E_B|&\le\tau_{\max}\sqrt{kL/2},\\
\left|\sum_{j\in B}(Z_j-\pi)(\mu_j-\bar\mu_B)\right|
&\le2\mu_{\max}\sqrt{kL/2}
\end{aligned}
\right\}.
\]

Conditional on the assignment, independence and sub-Gaussianity
of the errors give
\[
\mathbb E\left[
\exp\left\{t\sum_{j\in B}(Z_j-\bar Z_B)\epsilon_j\right\}
\,\middle|\,Z_{[n]}\right]
\le\exp\left\{\frac{\sigma^2t^2}{2}
\sum_{j\in B}(Z_j-\bar Z_B)^2\right\}
\le e^{\sigma^2t^2k/2}.
\]
By the sub-Gaussian tail bound and union bounds, the following
events fail with probabilities at most $2n^2e^{-L}$ and
$2ne^{-L}$, respectively:
$$
\mathcal E_2
:=
\bigcap_{B\in\mathcal N}
\left\{
\left|\sum_{j\in B}(Z_j-\bar Z_B)\epsilon_j\right|
\le\sigma\sqrt{2kL}
\right\},
$$
$$
\mathcal E_3
:=
\left\{
\max_{j\in[n]}|\epsilon_j|\le\sigma\sqrt{2L}
\right\}.
$$
These statements also hold when $\sigma=0$.

Here, $\mathcal E_1$ controls the deviations caused by treatment
assignment, $\mathcal E_2$ controls the error contribution to
each neighbourhood's statistic, and $\mathcal E_3$ bounds every
individual error.
Let $\mathcal E:=\mathcal E_1\cap\mathcal E_2\cap\mathcal E_3$.
By the union bound,
\begin{equation}
\begin{aligned}
\mathbb P(\mathcal E^c)
&=\mathbb P(\mathcal E_1^c\cup\mathcal E_2^c\cup\mathcal E_3^c)\\
&\le
\mathbb P(\mathcal E_1^c)
+\mathbb P(\mathcal E_2^c)
+\mathbb P(\mathcal E_3^c)\\
&\le\{6n^2(n+1)+2n^2+2n\}e^{-L}\\
&\le16n^3e^{-L}=\zeta/2,
\end{aligned}
\label{eq:knn_good_event_proof}
\end{equation}
where the last line uses $n\ge2$ and
$L=\log(32n^3/\zeta)$.

\paragraph{2. Exact randomization tests.}
For $B\in\mathcal N$, write
$\bar Y_B=k^{-1}\sum_{j\in B}Y_j$ and
$T_B(z)=\sum_{j\in B}(z_j-\pi)(Y_j-\bar Y_B)$.
Since $Y_j=\mu_j+Z_j\tau_j+\epsilon_j$,
\begin{align*}
T_B(Z)
=\sum_{j\in B}(Z_j-\pi)(\mu_j-\bar\mu_B)
+\sum_{j\in B}(Z_j-\bar Z_B)\epsilon_j
+(1-\bar Z_B)\sum_{j\in B}Z_j\tau_j,
\end{align*}
where
\[
(1-\bar Z_B)\sum_{j\in B}Z_j\tau_j
=\pi(1-\pi)A_B+(1-\bar Z_B)E_B-\frac{\pi A_B}{k}D_B.
\]
Using $0\le\bar Z_B\le1$, $A_B/k\le\tau_{\max}$, and the
bounds on $\mathcal E$, we obtain
\begin{equation}
T_B(Z)\ge\kappa A_B-\sqrt2\,s_Y\sqrt{kL},
\qquad B\in\mathcal N.
\label{eq:knn_observed_lower_bound_proof}
\end{equation}

Let $\mathcal D=(Y,Z)$ denote the observed outcomes and
assignments. Conditional on $\mathcal D$, complete
randomization gives $\operatorname{Var}(Z'_j)=\pi(1-\pi)$ and
$\operatorname{Cov}(Z'_j,Z'_\ell)=-\pi(1-\pi)/(n-1)$ for
$j\ne\ell$. Since the centered coefficients sum to zero,
\begin{equation}
\begin{aligned}
\mathbb E\{T_B(Z')\mid\mathcal D\}&=0,\\
\operatorname{Var}\{T_B(Z')\mid\mathcal D\}
&=\frac{n\pi(1-\pi)}{n-1}
\sum_{j\in B}(Y_j-\bar Y_B)^2.
\end{aligned}
\label{eq:knn_reference_variance_proof}
\end{equation}
On $\mathcal E$, $|Y_j|\le\sqrt{2L}\,s_Y$. Therefore
\[
\operatorname{Var}\{T_B(Z')\mid\mathcal D\}
\le\frac12\sum_{j\in B}Y_j^2\le s_Y^2kL.
\]
Set $C_0=(\sqrt2+2/\sqrt\alpha)/\kappa$. Whenever
\begin{equation}
A_B>C_0s_Y\sqrt{kL},
\label{eq:knn_regional_signal_proof}
\end{equation}
the observed statistic $T_B(Z)$ exceeds $2s_Y\sqrt{kL/\alpha}$. Then by 
Chebyshev's inequality and \eqref{eq:knn_reference_variance_proof}, simultaneously on $\mathcal E$,
\[
\begin{aligned}
p_B
:=\mathbb P\{T_B(Z')\ge T_B(Z)\mid\mathcal D\}
\le\frac{\operatorname{Var}\{T_B(Z')\mid\mathcal D\}}
{T_B(Z)^2}
<\frac{s_Y^2kL}{\left(2s_Y\sqrt{kL/\alpha}\right)^2}
=\frac{\alpha}{4}.
\end{aligned}
\]
\paragraph{3. Monte-Carlo tests.}
An explicit sufficient choice is
\begin{equation}
M\ge\frac{12}{\alpha}\log\left(\frac{2n^2}{\zeta}\right).
\label{eq:knn_mc_sufficient_proof}
\end{equation}
For $M$ independent randomized assignments, let $C_B$ count the
randomized statistics that are at least as large as $T_B(Z)$.
Thus, conditional on $\mathcal D$, $C_B\sim\operatorname{Binomial}(M,p_B)$, and we can write the Monte-Carlo $p$-value as
\[
\tilde p_B=(1+C_B)/(M+1). 
\]
Since $M\alpha\ge2$,
\[
\tilde p_B>\alpha
\quad\Longrightarrow\quad
C_B>\alpha(M+1)-1\ge M\alpha/2.
\]
For $p_B\le\alpha/4$, a Chernoff bound with $t=\log 2$ gives
\begin{align*}
\mathbb P\{\tilde p_B>\alpha\mid\mathcal D\}
&\le 2^{-M\alpha/2}\mathbb E(2^{C_B}\mid\mathcal D)\\
&=2^{-M\alpha/2}(1+p_B)^M\\
&=\exp\left\{-\frac{M\alpha}{2}\log 2\right\}(1+p_B)^M
\qquad (2^x=e^{x\log 2})\\
&\le\exp\left\{-\frac{M\alpha}{2}\log 2+Mp_B\right\}
\qquad (1+x\le e^x)\\
&\le\exp\left\{-\frac{M\alpha}{2}\log 2+\frac{M\alpha}{4}\right\}
\qquad (p_B\le\alpha/4)\\
&\le e^{-M\alpha/12}.
\end{align*}
With $\mathcal E$ as defined above, let
\[
\mathcal E_*
:=
\mathcal E
\cap
\bigcap_{\substack{B\in\mathcal N\\
A_B>C_0s_Y\sqrt{|B|L}}}
\{\tilde p_B\le\alpha\}.
\]
Thus, $\mathcal E_*$ is the event that all concentration
bounds defining $\mathcal E$ hold and every neighbourhood
satisfying \eqref{eq:knn_regional_signal_proof} is rejected.
On $\mathcal E$, the preceding Monte Carlo bound applies
to each such neighbourhood. Combining
\eqref{eq:knn_good_event_proof} with a union bound over
at most $n^2$ neighbourhoods gives
\[
\mathbb P(\mathcal E_*)
\ge1-\zeta/2-n^2e^{-M\alpha/12}
\ge1-\zeta.
\]

\paragraph{4. The excess-size bound.}
For $u=(k-K_0)/n>0$, the geometry bound gives
\begin{equation}
\label{equ:Aik}
\frac{A_{i,k}}{s_Y\sqrt{kL}}
\ge\frac{\lambda\sqrt n}{s_Y\sqrt L}
   \frac{u^\gamma}{\sqrt{K_0/n+u}}.
\end{equation}
Define
\[
u_*:=\max\left\{
\left(\frac{4C_0^2s_Y^2K_0L}{\lambda^2n^2}\right)^{1/(2\gamma)},
\left(\frac{4C_0^2s_Y^2L}{\lambda^2n}\right)^{1/(2\gamma-1)}
\right\}>0.
\]
Then
\begin{equation}
u_*^{2\gamma}\ge\frac{4C_0^2s_Y^2L}{\lambda^2n}\frac{K_0}{n},
\qquad
u_*^{2\gamma}\ge\frac{4C_0^2s_Y^2L}{\lambda^2n}u_*.
\label{eq:knn_ustar_bounds_proof}
\end{equation}
Adding the two inequalities in \eqref{eq:knn_ustar_bounds_proof},
dividing by two, and then taking square roots and rearranging gives
\begin{equation}
\label{equ:lambda_n_sy}
\frac{\lambda\sqrt n}{s_Y\sqrt L}
\frac{u_*^\gamma}{\sqrt{K_0/n+u_*}}
\ge\sqrt2C_0>C_0.
\end{equation}
Moreover,
\[
\frac{d}{du}\log\left(\frac{u^\gamma}{\sqrt{K_0/n+u}}\right)
=\frac{2\gamma K_0/n+(2\gamma-1)u}{2u(K_0/n+u)}>0.
\]
The positive derivative shows that
$u^\gamma/\sqrt{K_0/n+u}$ is increasing for $u>0$.
Therefore, whenever $u=(k-K_0)/n\ge u_*$, the 
bounds in \eqref{equ:Aik} and \eqref{equ:lambda_n_sy} gives, for every centre $i$,
\[
\frac{A_{i,k}}{s_Y\sqrt{kL}}
\ge
\frac{\lambda\sqrt n}{s_Y\sqrt L}
\frac{u_*^\gamma}{\sqrt{K_0/n+u_*}}
\ge\sqrt2C_0>C_0.
\]
Thus $A_{i,k}>C_0s_Y\sqrt{kL}$, which is
the condition in \eqref{eq:knn_regional_signal_proof}
for the neighbourhood $N_i^{(k)}$.

Put $h_*=\lceil nu_*\rceil$. For every $k\in[n]$ with
$k\ge K_0+h_*$,
\[
\frac{k-K_0}{n}\ge\frac{h_*}{n}\ge u_*.
\]
Every neighbourhood at such a size therefore satisfies
\eqref{eq:knn_regional_signal_proof}. By the definition
of $\mathcal E_*$, all these neighbourhoods are rejected
on $\mathcal E_*$, giving
\[
\tilde p_k=\max_{i\in[n]}\tilde p_{i,k}\le\alpha.
\]
If $K_0+h_*\le n$, the largest non-rejected size therefore
satisfies $\widehat K_0^{\,U}\le K_0+h_*-1$.
If $K_0+h_*>n$, the same inequality follows from
$\widehat K_0^{\,U}\le n\le K_0+h_*-1$, since
$K_0+h_*$ and $n$ are integers.
Thus, in both cases, on $\mathcal E_*$,
\[
\frac{\widehat K_0^{\,U}-K_0}{n}
\le\frac{h_*-1}{n}<u_*,
\]
by $h_*-1<nu_*$.
By the definition of $u_*$, since $\gamma\ge1$ and $4C_0^2>1$,
\[
u_*
\le4C_0^2\left\{
\left(\frac{s_Y^2K_0L}{\lambda^2n^2}\right)^{1/(2\gamma)}
+\left(\frac{s_Y^2L}{\lambda^2n}\right)^{1/(2\gamma-1)}
\right\}.
\]
This proves \eqref{eq:knn_power_rate}, with an implicit constant
depending only on $\pi_{\min}$ and $\alpha$. The probability
includes the errors and the observed and simulated assignments.

For the simplified asymptotic rate, hold $\zeta$, $s_Y/\lambda$,
and $\gamma$ fixed. Then $L=O(\log n)$ and $K_0\le n$, so the
first term is $O\{(\log n/n)^{1/(2\gamma)}\}$. The other terms
are of no larger order because $\gamma\ge1$. Substituting
$\gamma=\nu+\beta$ gives the final display.
\end{proof}

\subsection{Proof of Theorem~\ref{thm:knn_minimax_lower}}
\label{app:proof_knn_minimax_lower}

\paragraph{Setup.}
Fix $d\ge1$, $\sigma>0$, $\gamma\ge1$, and
$0<\lambda\le\tau_{\max}/(16\gamma)$. For this lower-bound
argument, use the normalized ranks $v_j$ as the fixed covariates
\[
W_j=v_j=(j/n,\dots,j/n),\qquad j\in[n],
\]
and the Gaussian outcome model
\[
Y_j(0)=\epsilon_j,\qquad
Y_j(1)=\epsilon_j+\tau_j,\qquad
\epsilon_j\stackrel{\mathrm{i.i.d.}}{\sim}N(0,\sigma^2).
\]
This is a special case of Assumption~\ref{ass:knn_power_model},
used only to prove the lower bound. 

Treatment follows the same
complete-randomization distribution under every model and is
independent of the errors. Write $\mathbb P_\tau$ for the law
of the observed data, with $Y_j=Z_j\tau_j+\epsilon_j$, and
any auxiliary randomization used to compute $U$. The conditional
law of this auxiliary randomization given the data does not
depend on $\tau$. All probabilities condition on the fixed
covariates.

Since $\|v_i-v_j\|_2=\sqrt d\,|i-j|/n$, each nearest-neighbour
set is a contiguous block. Conversely, with distance ties
broken in favour of the smaller rank,
\[
\{s,\dots,s+k-1\}
=N_{s+\lfloor k/2\rfloor}^{(k)},
\qquad 1\le s\le n-k+1.
\]
Thus the candidate family consists of all contiguous blocks,
and $K_0(\tau)$ is the longest run of zero entries in $\tau$.

For this design, $\mathcal F_n(\lambda,\gamma)$ contains all
$\tau\in[0,\tau_{\max}]^n$ satisfying
Assumption~\ref{ass:knn_geometry} for some nonempty closed
zero-effect set and parameters with
$\lambda_\tau ca^\beta=\lambda$ and $\nu+\beta=\gamma$.

Our construction uses $\beta=0$ and $\nu=\gamma$.
In this case, the assumption requires a nonempty closed set
$\mathcal Z_0\subseteq[0,1]^d$ and constants
$\lambda_\tau,c>0$ with $\lambda_\tau c=\lambda$ such that
\[
\begin{cases}
\tau_j=0, & v_j\in\mathcal Z_0,\\
\lambda_\tau\le\tau_j\le\tau_{\max},
& v_j\notin\mathcal Z_0,
\end{cases}
\qquad j\in[n],
\]
and, for every contiguous block $I$ with $|I|>K_0$,
\[
\#\{j\in I:\tau_j>0\}
\ge cn\left(\frac{|I|-K_0}{n}\right)^\gamma.
\]
We verify a lower bound on each positive effect
and on the number of affected units in every block larger
than $K_0$. No condition on their distances from
$\mathcal Z_0$ is required.

\begin{proof}
\textbf{Roadmap.}
Let $U$ satisfy the coverage condition in the theorem.
We first construct a reference model and alternatives whose
targets differ by $h$, then verify that they belong to
$\mathcal F_n(\lambda,\gamma)$. We next show that a mixture
of the alternative distributions is close to the reference
distribution. Coverage under the reference model then forces
$U-K_0\ge h$ with probability greater than $\zeta$ under
at least one alternative.

\paragraph{1. Construction of the models.}
Write $\Delta:=\tau_{\max}$ and $c_\gamma:=1/(16\gamma)$.
Let $m=\lfloor n/4\rfloor$ and choose an integer
$1\le h\le m/2$; we specify $h$ in part~4.
We construct a reference model with $K_0=m+h$ and
alternatives with $K_0=m$. Define the blocks
\[
A:=\{1,\dots,m\},\qquad
D:=\{m+1,\dots,2m\},\qquad
B:=\{2m+1,\dots,3m+h\}.
\]
Use the superscript $(0)$ to label the reference model.
Its treatment-effect vector and data distribution are
\[
\tau_j^{(0)}:=\Delta\one\{j\notin A\cup B\},
\qquad \mathbb P_0:=\mathbb P_{\tau^{(0)}}.
\]
Thus, units in $A\cup B$ have zero effects, and all other
units have effect $\Delta$.
The zero-effect blocks $A$ and $B$ have lengths $m$ and
$m+h$, separated by the affected block $D$. Hence
$K_0(\tau^{(0)})=m+h$.

We obtain alternatives by adding positive effects inside $B$.
To ensure that its remaining zero-effect blocks have length
at most $m$, choose the added units from
\[
J:=\{2m+h+1,\dots,3m\},\qquad
\ell:=|J|=m-h,
\]
and let their number be
\[
s:=\left\lceil c_\gamma n(h/n)^\gamma\right\rceil.
\]
Since $\gamma\ge1$ and $h\le m/2$,
\[
0<c_\gamma n(h/n)^\gamma
=c_\gamma h(h/n)^{\gamma-1}\le h,
\qquad 1\le s\le h\le\ell.
\]
Each alternative is indexed by a subset $S\subseteq J$
with $|S|=s$. The set $S$ identifies the units whose effects
we change from zero to $\Delta$; all other effects remain
as in the reference model. Write $\tau^{(S)}$ for the
resulting treatment-effect vector and $\mathbb P_S$ for
its data distribution:
\[
\tau_j^{(S)}:=\tau_j^{(0)}+\Delta\one\{j\in S\},
\qquad \mathbb P_S:=\mathbb P_{\tau^{(S)}}.
\]
Every subinterval of $B$ with more than $m$ units starts
at or before $2m+h$ and ends at or after $3m+1$.
It therefore contains $J$, and hence all the affected units
in $S$. No zero-effect block can cross $D$ or the affected
units outside $A\cup B$. Since $A$ remains unaffected,
\begin{equation}
K_0(\tau^{(S)})=m,
\qquad
K_0(\tau^{(0)})-K_0(\tau^{(S)})=h.
\label{eq:minimax_target_separation}
\end{equation}

\paragraph{2. Verification of the assumptions.}
All positive effects equal $\Delta$. We now check that every
block larger than $K_0$ contains enough affected units.
For a block $I$, let
$n_+(I):=\#\{j\in I:\tau_j>0\}$ under the model being
considered, and define
\[
F(t):=c_\gamma n(t/n)^\gamma,\qquad 0\le t\le n.
\]
For $0<t\le n$, we have
\[
0\le F(t)\le c_\gamma t\le t,
\qquad
0\le F'(t)=c_\gamma\gamma(t/n)^{\gamma-1}\le1/16.
\]
We will show that every constructed model satisfies
\begin{equation}
n_+(I)\ge F(|I|-K_0)
=c_\gamma n\left(\frac{|I|-K_0}{n}\right)^\gamma
\quad\text{whenever }|I|>K_0.
\label{eq:minimax_count_verification}
\end{equation}

First consider an alternative $\tau^{(S)}$. For a block $I$
of length $k>m$, put $t=k-m$. There are three cases.
If $I$ meets both $A$ and $B$, it contains all of $D$, so
for $n\ge8$,
\[
n_+(I)\ge m=\lfloor n/4\rfloor
\ge n/8\ge c_\gamma n\ge F(t).
\]
If $I$ meets at most one of $A,B$ and $|I\cap B|\le m$,
it contains at most $m$ unaffected units, giving
\[
n_+(I)\ge k-m=t\ge F(t).
\]
In the remaining case, $I$ meets $B$ but not $A$ and
$|I\cap B|>m$. Part~1 shows that $I\cap B$ contains $J$,
so $I$ contains all $s$ affected units in $S$. Every unit
of $I$ outside $B$ is also affected. Writing
$x_+:=\max\{x,0\}$, we obtain
\[
n_+(I)\ge s+\{k-(m+h)\}_+
\ge F(h)+(t-h)_+\ge F(t).
\]
For $t\le h$, the last inequality follows from
$F(h)\ge F(t)$. For $t>h$, it follows from
\[
F(t)-F(h)=\int_h^t F'(u)\,du\le t-h.
\]
This proves \eqref{eq:minimax_count_verification} for
every alternative.

For the reference model, take $k>m+h$ and set $t=k-(m+h)$.
If $I$ meets at most one of $A,B$, it contains at most
$m+h$ unaffected units, so $n_+(I)\ge t\ge F(t)$.
If it meets both, it contains $D$ and
$n_+(I)\ge m\ge F(t)$ as above. Thus
\eqref{eq:minimax_count_verification} also holds for the
reference model.

For each constructed vector $\tau$, choose
\[
\mathcal Z_0:=\{v_j:\tau_j=0\},\qquad
\lambda_\tau=\Delta,\qquad a=1,\qquad
c=\lambda/\Delta,\qquad \beta=0,\qquad \nu=\gamma.
\]
The set $\mathcal Z_0$ is nonempty and finite, hence closed,
and contains exactly the sampled zero-effect points.
Every positive effect equals $\Delta=\lambda_\tau$, and
\[
c=\lambda/\Delta\le c_\gamma,\qquad
\lambda_\tau ca^\beta=\lambda,\qquad \nu+\beta=\gamma.
\]
Consequently, \eqref{eq:minimax_count_verification} implies
the required count bound with constant $c$. These parameters
do not depend on $n$, and
\begin{equation}
\tau^{(0)},\tau^{(S)}\in\mathcal F_n(\lambda,\gamma)
\qquad\text{for every }S.
\label{eq:minimax_class_membership}
\end{equation}

\paragraph{3. Distance between the distributions.}
Choose $S$ uniformly from the $\binom{\ell}{s}$ subsets
of $J$ containing exactly $s$ units, and then generate
the data under the corresponding alternative model
$\mathbb P_S$. Denote the resulting distribution by
$\overline{\mathbb P}$. Thus, for any measurable event $E$,
\[
\overline{\mathbb P}(E)
:=\frac{1}{\binom{\ell}{s}}
\sum_{\substack{S\subseteq J\\|S|=s}}\mathbb P_S(E).
\]
We compare this mixture with the reference distribution
$\mathbb P_0$.
For probability laws $P,Q$, their total variation distance is
$\operatorname{TV}(P,Q):=\sup_E|P(E)-Q(E)|$, with the
supremum over measurable events $E$. We bound this distance to compare the probability of
$\{U\ge m+h\}$ under the reference model and the mixture
in part~4.

The assignment law is the same under both models. Conditional
on $Z_{[n]}$, the outcomes are independent, and for $j\in S$,
\[
Y_j\sim
\begin{cases}
N(0,\sigma^2), & \text{under }\mathbb P_0,\\
N(Z_j\Delta,\sigma^2), & \text{under }\mathbb P_S.
\end{cases}
\]
The outcome distributions outside $S$ are identical under
the two models. Thus, the assignment probabilities and the
outcome-density factors outside $S$ cancel, giving
\[
\begin{aligned}
L_S
:=\frac{d\mathbb P_S}{d\mathbb P_0}
&=\prod_{j\in S}
\frac{\exp\{-(Y_j-Z_j\Delta)^2/(2\sigma^2)\}}
     {\exp\{-Y_j^2/(2\sigma^2)\}}\\
&=\exp\left\{
\frac{1}{2\sigma^2}\sum_{j\in S}
\bigl[Y_j^2-(Y_j-Z_j\Delta)^2\bigr]
\right\}\\
&=\exp\left\{
\frac{1}{2\sigma^2}\sum_{j\in S}
(2\Delta Z_jY_j-\Delta^2Z_j^2)
\right\}\\
&=\exp\left\{
\frac{\Delta}{\sigma^2}\sum_{j\in S}Z_jY_j
-\frac{\Delta^2}{2\sigma^2}\sum_{j\in S}Z_j
\right\},
\end{aligned}
\]
where the last equality uses $Z_j^2=Z_j$.
Under $\mathbb P_0$, conditional on $Z_{[n]}$, the outcomes
on $J$ are independent $N(0,\sigma^2)$. Write
$\mathbb E_0$ for expectation under $\mathbb P_0$ and
$b:=\Delta^2/\sigma^2$. Using the Gaussian identity
$\mathbb E e^{t\epsilon_j}=e^{\sigma^2t^2/2}$ gives
\[
\mathbb E_0\{L_SL_{S'}\mid Z_{[n]}\}
=\prod_{j\in S\cap S'}e^{bZ_j}
=\exp\left\{b\sum_{j\in S\cap S'}Z_j\right\}.
\]
A unit appearing in only one of the two sets contributes
expectation one; a unit in both contributes $e^{bZ_j}$.
No independence between treatment indicators is required.

Let $S,S'$ be independent uniform subsets of size $s$ in $J$,
independent of the data. The mixture likelihood ratio is
\[
\bar L:=\frac{d\overline{\mathbb P}}{d\mathbb P_0}
=\mathbb E_S L_S.
\]
Write $\mathbb E_Z$ for expectation over the common assignment
law. Expanding the square and using $Z_j\le1$, we obtain
\[
\begin{aligned}
\mathbb E_0\bar L^2
&=\mathbb E_{S,S'}\mathbb E_0(L_SL_{S'})\\
&=\mathbb E_{S,S'}\mathbb E_Z
  \exp\left\{b\sum_{j\in S\cap S'}Z_j\right\}\\
&\le\mathbb E_{S,S'}e^{b|S\cap S'|}.
\end{aligned}
\]
To bound the overlap, put $H:=|S\cap S'|$. For fixed $S$,
each of its $\binom sr$ subsets of size $r$ is contained
in $S'$ with probability
$\binom{\ell-r}{s-r}/\binom\ell s$. Therefore, for
$r=0,\dots,s$,
\[
\mathbb E\binom Hr
=\binom sr\frac{\binom{\ell-r}{s-r}}{\binom\ell s}
=\binom sr\prod_{q=0}^{r-1}\frac{s-q}{\ell-q}
\le\binom sr(s/\ell)^r.
\]
The empty product for $r=0$ equals one, and the inequality
uses $(s-q)/(\ell-q)\le s/\ell$. By the binomial theorem,
\[
\begin{aligned}
\mathbb E e^{bH}
&=\mathbb E\{1+(e^b-1)\}^H\\
&=\sum_{r=0}^s(e^b-1)^r\mathbb E\binom Hr\\
&\le\sum_{r=0}^s\binom sr
  \left\{\frac{s}{\ell}(e^b-1)\right\}^r\\
&=\left\{1+\frac{s}{\ell}(e^b-1)\right\}^{s}
\le\exp\left\{(e^b-1)\frac{s^2}{\ell}\right\},
\end{aligned}
\]
where the last inequality uses $1+x\le e^x$.
Since $\mathbb E_0\bar L=1$, Cauchy--Schwarz now gives
\begin{equation}
\begin{aligned}
\operatorname{TV}(\overline{\mathbb P},\mathbb P_0)
&=\frac12\mathbb E_0|\bar L-1|\\
&\le\frac12\sqrt{\mathbb E_0(\bar L-1)^2}
=\frac12\sqrt{\mathbb E_0\bar L^2-1}\\
&\le\frac12\left[
\exp\left\{(e^b-1)\frac{s^2}{\ell}\right\}-1
\right]^{1/2}.
\end{aligned}
\label{eq:knn_minimax_tv}
\end{equation}
Adding the auxiliary randomization cannot increase this
distance, because its conditional law given the data is
the same under all models. Thus the bound also applies
to events involving $U$.

\paragraph{4. Lower bound on the excess.}
Set $\eta:=(1-\alpha-\zeta)/2>0$ and choose
\[
h=\left\lfloor b_0n^{1-1/(2\gamma)}\right\rfloor
\]
for a sufficiently small constant $b_0>0$. Since
$h\to\infty$ and $h/n\to0$, we have $1\le h\le m/2$
for all sufficiently large $n$. Recall from part~1 that
\[
s=\left\lceil c_\gamma n(h/n)^\gamma\right\rceil,
\qquad
\ell=m-h,
\qquad
m=\lfloor n/4\rfloor.
\]
With $h=\lfloor b_0n^{1-1/(2\gamma)}\rfloor$ chosen above,
these definitions give
\[
\frac{s}{\sqrt n}\longrightarrow c_\gamma b_0^\gamma,
\qquad
\frac{\ell}{n}\longrightarrow\frac14,
\qquad
\frac{s^2}{\ell}
=\frac{(s/\sqrt n)^2}{\ell/n}
\longrightarrow4c_\gamma^2b_0^{2\gamma}.
\]
Choose $b_0$ small enough that
\[
\frac12\left[
\exp\{4(e^b-1)c_\gamma^2b_0^{2\gamma}\}-1
\right]^{1/2}<\eta.
\]
Then \eqref{eq:knn_minimax_tv} gives
$\operatorname{TV}(\overline{\mathbb P},\mathbb P_0)<\eta$
for sufficiently large $n$.

The reference vector belongs to $\mathcal F_n(\lambda,\gamma)$
by \eqref{eq:minimax_class_membership}, so the coverage
condition gives $\mathbb P_0\{U\ge m+h\}\ge1-\alpha$.
By the definition of total variation,
\[
\begin{aligned}
\overline{\mathbb P}\{U\ge m+h\}
&\ge\mathbb P_0\{U\ge m+h\}
  -\operatorname{TV}(\overline{\mathbb P},\mathbb P_0)\\
&>1-\alpha-\eta
=\frac{1-\alpha+\zeta}{2}>\zeta.
\end{aligned}
\]
The mixture probability is the average of the probabilities
under the alternatives. Since this average exceeds $\zeta$,
at least one $S$ satisfies $\mathbb P_S\{U\ge m+h\}>\zeta$.
For this alternative, \eqref{eq:minimax_target_separation}
gives $K_0(\tau^{(S)})=m$, and therefore
\[
\mathbb P_S\{U-K_0(\tau^{(S)})\ge h\}>\zeta.
\]
For sufficiently large $n$,
$h/n\ge(b_0/2)n^{-1/(2\gamma)}$. Taking $c_*:=b_0/2$ yields
\[
\sup_{\tau\in\mathcal F_n(\lambda,\gamma)}
\mathbb P_\tau\left\{
\frac{U-K_0(\tau)}{n}\ge c_*n^{-1/(2\gamma)}
\right\}>\zeta.
\]
This proves the theorem. The constant $c_*$ depends only on
$\alpha,\zeta,\gamma,\sigma$, and $\tau_{\max}$, and is
independent of $n$, $U$, and the dimension $d$.
\end{proof}

\section{Supporting calculations}
\label{app:supporting-calculations}

\subsection{Equivalence with contiguous blocks in one dimension}
\label{app:one_dimensional_neighbour_equivalence}

We prove \eqref{eq:1d_neighbour_blocks}. Index the units by their
covariate ranks and use distance $|i-j|/n$, breaking distance ties
in favour of the smaller rank. Fix a centre $i$ and a size $r$.
Its nearest-neighbour set contains $i$. If this set contained two
ranks with a missing rank between them, that missing rank would be
closer to $i$ than a selected rank on the same side of $i$.
This contradicts the definition of the nearest neighbours. Thus every
$N_i^{(r)}$ is a contiguous block of length $r$.

Conversely, consider $B_s^{(r)}=\{s,\ldots,s+r-1\}$ and put
$c=s+\lfloor r/2\rfloor$. If $r=2q+1$, the block consists of
the centre $c$ and the $q$ ranks on each side. All ranks outside
the block have strictly greater distance from $c$. If $r=2q$,
the block is $\{c-q,\ldots,c+q-1\}$. Its two possible boundary
candidates at distance $q/n$ are $c-q$ and $c+q$; the tie rule
selects $c-q$. The other $2q-1$ selected ranks have smaller
distance. This argument also applies when one of the outside
candidates does not exist at the sample boundary. Hence
$N_c^{(r)}=B_s^{(r)}$ in both cases, proving
\[
\{N_i^{(r)}:i\in[n]\}=\mathcal B_r.
\]
Repeated sets do not change a maximum. Therefore the local-null
families, maximum $p$-values, largest zero-effect sizes, and inverted
bounds agree. For Monte Carlo tests, the equality is exact when both
formulations use the same shared assignments: the same subset then
has the same observed and randomized statistics.

\subsection{Derivation of the one-dimensional computational costs}
\label{app:one_dimensional_computation_derivation}

We justify Table~\ref{tab:computational_costs}. These costs start
from the common covariate ordering; forming this ordering costs an
additional $O(n\log n)$ time. The implementation of conditional
tests considered here generates assignments separately for each
block and sums its statistic from scratch, at cost $O(Mr)$ for a
block of length $r$. There are
$n-r+1$ such blocks. Therefore the full conditional search costs
\[
O\!\left(M\sum_{r=1}^n r(n-r+1)\right)
=O\!\left(M\frac{n(n+1)(n+2)}6\right)
=O(Mn^3).
\]
For a descending search checking the $t$ largest lengths, write
$s=n-r+1$. The corresponding cost is
\[
O\!\left(M\sum_{s=1}^t s(n-s+1)\right)
=O\!\left(M\frac{t(t+1)(3n+2-2t)}6\right)
=O(Mnt^2).
\]
These costs describe the implementation just specified. They are
not lower bounds for every implementation of conditional tests.

For shared tests, form the global cumulative sums of the outcomes,
treatment indicators, and their products in $O(Mn)$ time, including
the observed assignment. The differences between cumulative sums in
\eqref{eq:block_stat_cumulative} give each block statistic in constant
time per assignment. At a fixed length $r$, all statistics and
Monte Carlo comparisons therefore cost $O\{M(n-r+1)\}$ time.
Since $\sum_{r=1}^n(n-r+1)=n(n+1)/2$, the full search costs
$O(Mn^2)$. Checking $t$ arbitrary lengths costs at most $O(Mnt)$,
including preprocessing. For the $t$ largest lengths, the sharper
count is $O(Mn+Mt^2)$, which is also $O(Mnt)$ because $1\le t\le n$.

Under \eqref{eq:mc_monotonicity}, non-rejection occurs at all
lengths up to some value, and rejection occurs at every larger length.
After checking the endpoints, bisection halves the remaining interval
at each step and finds the largest non-rejected length by testing
$O(\log n)$ lengths. Testing all blocks at one length costs at most $O(Mn)$,
giving $O(Mn\log n)$ time. This algorithm returns exactly the
largest non-rejected length.

The Gaussian approximation uses one additional cumulative sum of
$Y_i^2$. The variance identity established below gives all Gaussian
block $p$-values at a length in $O(n)$ time. Evaluating
$O(\log n)$ lengths therefore costs $O(n\log n)$. If the
subsequent Monte Carlo checks and bracket refinements use $k$
length evaluations, their cost is $O(Mnk)$, for a total of
$O(n\log n+Mnk)$. Here $k$ counts every Monte Carlo evaluation,
including endpoint checks. If two adjacent endpoints satisfy
\eqref{eq:mc_bracket_conditions}, monotonicity makes those two
checks sufficient, giving $k=2$. No accuracy guarantee for the
Gaussian approximation is needed for this accounting: the final
bracket is checked using the Monte Carlo tests.

The bisection costs rely on the one-dimensional common ordering.
For nearest-neighbour regions in higher dimensions, evaluating the
tests at selected sizes using this implementation requires cumulative
sums for each centre. Constructing all of those sums already costs
$O(Mn^2)$, so testing fewer sizes does not by itself give the same
computational saving.

\subsection{Cost of an exhaustive rectangle search}
\label{app:full_rank_rectangle_cost}

An ordered coordinate has $n(n+1)/2$ contiguous rank intervals.
Taking one interval in each of two coordinates gives
\[
\left\{\frac{n(n+1)}2\right\}^2=O(n^4)
\]
grid rectangles. Distinct rectangles may contain the same units,
and some rectangles may be empty; the count describes the
exhaustive grid enumeration and does not require the induced
unit sets to be distinct.

On the $n\times n$ rank grid, accumulate cell counts, outcome sums,
treatment sums, and treatment--outcome product sums. Counts and
outcome sums are fixed across assignments. For each assignment,
the corresponding two-dimensional cumulative sums can be formed in
$O(n^2)$ time, including zero entries for empty cells. Each rectangle
sum is then obtained from four cumulative-sum entries. For a
nonempty rectangle, its mean, centred statistic, and Monte Carlo
comparison take constant time. Empty rectangles are skipped. Exhaustively
testing all enumerated rectangles therefore costs $O(Mn^4)$ time.
This is an algorithmic upper bound for the full enumeration; it does
not assert that every algorithm for rectangular regions must have
quartic cost.

\subsection{Exact variance of the shared randomized statistic}
\label{app:shared_reference_variance_derivation}

Suppose $n\ge2$ and treatment is assigned by complete randomization
with $0<n_1<n$ and $\pi=n_1/n$. Hold the observed outcomes fixed,
as required when computing the reference distribution. For a
nonempty region $B$, define
$a_i=\one\{i\in B\}(Y_i-\bar Y_B)$, so
$\sum_{i=1}^n a_i=0$ and $T_B(Z^{(m)})=\sum_i a_iZ_i^{(m)}$.
Under a fresh complete-randomization assignment,
\[
\mathbb E Z_i^{(m)}=\pi,\qquad
\operatorname{Var}(Z_i^{(m)})=\pi(1-\pi),\qquad
\operatorname{Cov}(Z_i^{(m)},Z_j^{(m)})
=-\frac{\pi(1-\pi)}{n-1}\quad(i\ne j).
\]
The covariance formula follows from
$\mathbb E(Z_i^{(m)}Z_j^{(m)})=n_1(n_1-1)/\{n(n-1)\}$.
Consequently the reference mean is zero and
\[
\begin{aligned}
\operatorname{Var}\{T_B(Z^{(m)})\}
&=\pi(1-\pi)\sum_i a_i^2
-\frac{\pi(1-\pi)}{n-1}\sum_{i\ne j}a_ia_j\\
&=\frac{n\pi(1-\pi)}{n-1}\sum_i a_i^2\\
&=\frac{n\pi(1-\pi)}{n-1}
\left\{\sum_{i\in B}Y_i^2
-\frac1{|B|}\left(\sum_{i\in B}Y_i\right)^2\right\}.
\end{aligned}
\]
The second equality uses
$\sum_{i\ne j}a_ia_j=(\sum_i a_i)^2-\sum_i a_i^2
=-\sum_i a_i^2$.
The factor involves the full sample size $n$, because the shared
assignment is sampled on the full experiment. The variance identity
is exact; approximating the reference distribution by a Gaussian is
a separate step. If the variance is zero, all outcomes within $B$
are equal, every centred statistic is zero, and its exact and
Monte Carlo upper-tail $p$-values equal one.

\subsection{The boundary-cap calculation in Appendix~\ref{app:knn_dimension}}
\label{app:proof-boundary-cap}

We derive \eqref{eq:knn_cap_mass} for the continuous
geometric example. This calculation motivates the exponent
$\nu=(d+1)/2$ in Assumption~\ref{ass:knn_geometry}.
Theorem~\ref{thm:knn_power} assumes that the corresponding
count condition holds for every sample neighbourhood
larger than $K_0$.

Translate and rotate coordinates so that the zero-effect half-space
is $\{x:x_1\le0\}$ and the ball centre is $(-r_0,0,\ldots,0)$,
where $r_0>0$ is fixed. Write a point in the affected cap as
$(x_1,x')$, with $0<x_1\le s$ and $x'\in\mathbb R^{d-1}$.
The enlarged ball condition is
\[
(x_1+r_0)^2+\|x'\|_2^2\le(r_0+s)^2.
\]
For $d\ge2$, its cross-section at depth $x_1$ is a
$(d-1)$-dimensional ball of squared radius
\[
(r_0+s)^2-(r_0+x_1)^2=(s-x_1)(2r_0+s+x_1).
\]
If $0<s\le r_0$, then
$2r_0\le2r_0+s+x_1\le4r_0$ throughout the cap.
Writing $\omega_{d-1}$ for the volume of the unit ball in
$\mathbb R^{d-1}$, its full volume is therefore
\[
\begin{aligned}
\operatorname{Vol}_d(B_s\setminus\mathcal Z_0)
&=\omega_{d-1}\int_0^s
 [(s-x_1)(2r_0+s+x_1)]^{(d-1)/2}\,dx_1\\
&\asymp\int_0^s(s-x_1)^{(d-1)/2}\,dx_1
 =\frac{2}{d+1}s^{(d+1)/2}.
\end{aligned}
\]
The constants may depend on $d$ and $r_0$, but not on $s$.
The distance to the half-space is $x_1$ when $x_1>0$.
Restricting the same integral to $s/2\le x_1\le s$ gives
\[
\operatorname{Vol}_d\{x\in B_s:
  \operatorname{dist}(x,\mathcal Z_0)\ge s/2\}
\asymp\int_{s/2}^s(s-x_1)^{(d-1)/2}\,dx_1
\asymp s^{(d+1)/2}.
\]
Thus the part of the cap at distance at least $s/2$ from
$\mathcal Z_0$ has the same order of volume as the full cap. In dimension one these two lengths are $s$ and
$s/2$, which gives the same formula with exponent $(d+1)/2=1$.

The full added band has volume
\[
\omega_d\{(r_0+s)^d-r_0^d\}\asymp s.
\]
For example, the mean value theorem writes the expression in
braces as $d(r_0+\theta s)^{d-1}s$ for some $\theta\in(0,1)$.
If the covariate density is between positive finite constants
on the balls, every displayed volume comparison also holds for
probability mass. It follows that $u_s=P_V(B_s\setminus B_0)
\asymp s$ and
\[
P_V\{v\in B_s:\operatorname{dist}(v,\mathcal Z_0)\ge s/2\}
\asymp u_s^{(d+1)/2},
\]
which is \eqref{eq:knn_cap_mass}. In particular, $s/2$ is bounded
below by a fixed positive multiple of $u_s$.

If the corresponding count bound holds uniformly over all sample
neighbourhoods larger than $K_0$, Assumption~\ref{ass:knn_geometry}
can be applied with $\nu=(d+1)/2$. Then
$2\gamma=d+1+2\beta$, and substituting this identity in
Theorem~\ref{thm:knn_power} gives the displayed rate
$(\log n/n)^{1/(d+1+2\beta)}$.
The continuous calculation by itself neither establishes this
uniform sample condition nor supplies a lower bound for that more
restricted geometric model.

\end{document}